\documentclass[prd,twocolumn, superscriptaddress, 10pt]{revtex4-2}
\usepackage{graphicx}
\usepackage{dcolumn}
\usepackage{bm}
\usepackage[utf8]{inputenc}
\usepackage[T1]{fontenc}
\usepackage{float}
\usepackage{subfigure}
\usepackage{amssymb,amsmath}
\usepackage{color}
\usepackage{hyperref}
\usepackage{listings}
\usepackage[linesnumbered,ruled,vlined]{algorithm2e} 
\usepackage{subcaption}
\usepackage{balance} 
\usepackage{ulem}

\begin{document}

\title{Variational quantum simulation of ground states and thermal states for lattice gauge theory with multi-objective optimization}
\author{Lang-Xing Cheng}
\affiliation{Key Laboratory of Atomic and Subatomic Structure and Quantum Control (Ministry of Education),  Guangdong Basic Research Center of Excellence for Structure and Fundamental Interactions of Matter, and  School of Physics, South China Normal University, Guangzhou 510006, China}

\author{Dan-Bo Zhang}
\email{dbzhang@m.scnu.edu.cn}
\affiliation{Key Laboratory of Atomic and Subatomic Structure and Quantum Control (Ministry of Education),  Guangdong Basic Research Center of Excellence for Structure and Fundamental Interactions of Matter, and  School of Physics, South China Normal University, Guangzhou 510006, China}
\affiliation{Guangdong Provincial Key Laboratory of Quantum Engineering and Quantum Materials,  Guangdong-Hong Kong Joint Laboratory of Quantum Matter, and Frontier Research Institute for Physics,\\  South China Normal University, Guangzhou 510006, China}

\begin{abstract}
Variational quantum algorithms provide feasible approaches for simulating quantum systems and are applied widely. For lattice gauge theory, however, variational quantum simulation faces a challenge as local gauge invariance enforces a constraint on the physical Hilbert space. In this paper, we incorporate multi-objective optimization for variational quantum simulation of lattice gauge theory at zero and finite temperatures. By setting energy or free energy of the system and penalty for enforcing the local gauge invariance as two objectives, the multi-objective optimization can self-adjust the proper weighting for two objectives and thus faithfully simulate the gauge theory in the physical Hilbert space. Specifically, we propose variational quantum eigensolver and  variational quantum thermalizer for preparing the ground states and thermal states of lattice gauge theory, respectively. We demonstrate the quantum algorithms for a $Z_2$ lattice gauge theory with spinless fermion in one dimension. With numeral simulations, the multi-objective optimization shows that minimizing energy~(free energy) and enforcing the local gauge invariance can be achieved simultaneously at zero temperature~(finite temperature). The multi-objective optimization suggests a feasible ingredient for quantum simulation of complicated physical systems on near-term quantum devices.

\end{abstract}

\maketitle

\section{Introduction}
Obtaining the ground state and thermal state of a quantum system is an essential topic in quantum many-body physics. Many algorithms can solve these problems such as imaginary time evolution\cite{Motta_2019,McArdle_2019} and Monte-Carlo simulation\cite{Ren2022TowardsTG}. However, as the system size increases, the required resources rise exponentially, and traditional methods will be inefficient. With the development of quantum hardware, variational quantum algorithm (VQA) has emerged as the leading strategy for quantum simulations on Noisy Intermediate-Scale Quantum (NISQ) computers\cite{Cerezo_2021,Huang_2023,article,Preskill_2018,Bharti_2022}. By expressing target quantum states with parameterized quantum circuits~\cite{Benedetti_2019,PRXQuantum.2.040337,Abbas_2021} and optimizing the parameters with minimization of energy or free energy, variational quantum algorithms are feasible for the simulation of static properties of generic quantum many-body systems, e.g, eigensolver~(VQE) for preparing eigenstates~\cite{Peruzzo2014,Fedorov2021VQEMA,Higgott2019variationalquantum,Nakanishi_2019} and variational quantum thermalizer(VQT) for generating thermal states~\cite{verdon2019quantumhamiltonianbasedmodelsvariational,Selisko_2024,Consiglio_2024,Xie_2022}.

Among various quantum systems, lattice gauge theory~(LGT) stands out as it describes physical systems respecting local gauge invariance, which is ubiquitous for both condensed matter physics~\cite{Ichinose_2014} and high-energy physics~\cite{PhysRevD.10.2445,RevModPhys.51.659,RevModPhys.55.775}. A quantum system described by lattice gauge theory shares the same difficulty for simulation as typical quantum many-body systems, although some classical methods, for instance, Monte-Carlo simulation~\cite{Yan2023,article2,PhysRevD.107.014505,article3} and tensor networks~\cite{Bauls2018TensorNA,Silvi_2014,PhysRevLett.112.201601}, can apply under some specific conditions. Likewise, variational quantum algorithms serve as a potential candidate for effectively simulating lattice gauge theories. Nevertheless, lattice gauge theories differ from other quantum systems in that the quantum states of LGT should satisfy Gauss law at each spatial site due to the local gauge invariance. 
The Gauss law can be enforced exactly by constructing a gauge-invariant parameterized quantum circuit~\cite{Mathew_2022,Mazzola_2021,Frank_2020}. However, the class of local gauge-invariant quantum gates is limited and thus restricts the expressive power of parameterized quantum circuits for solving complicated eigenstates or thermal states. Alternatively, the local gauge invariance can be satisfied approximately by punishing a violation of the Gauss law in the cost function or in the imaginary-time evolution~\cite{Davoudi-PRL-2023}. However, it is difficult to choose a proper weight for the penalty in the cost function~\cite{PhysRevResearch.3.013197,article5}. It is desirable for adaptively adjusting the strength of penalty to guarantee minimizing the energy or free energy in the physical Hilbert space. On the other hand, adjusting the weights among different learning tasks has been investigated intensively in multi-objective optimization~\cite{Caruana1997,9392366,articleq,NEURIPS2018_432aca3a}, which may sheds light on variational quantum simulation of lattice gauge theory. 

In this paper, we exploit multi-objective optimization for variational preparing ground states and thermal states of lattice gauge theory by quantum computing. By taking minimization of the energy or free energy and respecting the Gauss law as two objectives,  the multi-objective optimization allows self-adjusting the weighting between the two objectives and thus assures minimization of energy or free energy constraint in the physical Hilbert space. Using a one-dimensional lattice gauge theory with spinless fermion coupled to $Z_2$ gauge field as a demonstration, we show that variational quantum algorithms with multi-objective optimization can prepare ground states and thermal states of LGT with good accuracy. Our work showcases an application of multi-objective optimization for quantum simulation for LGT and the method may be generalized for other quantum problems with multiple objectives.

The paper is organized as follows. In Sec.~\ref{sec:one}, we introduce the basic concept  multi-objective optimization and how to incorporate it into variational quantum algorithms. In Sec.~\ref{sec:two}, we discuss the lattice gauge model and propose  variational approaches to obtain its ground state and thermal state. The simulation results are presented in Sec.~\ref{sec:three}. Finally, we give conclusions in Sec.~\ref{sec:four}.

\section{\label{sec:one}Multi-Objective Optimization}
In this section, we introduce multi-objective optimization as well as how to incorporate it into variational quantum algorithms. 

Multi-objective optimization arises in diverse fields where different objectives should be optimized instead of one. For instance, in machine learning, multi-task learning means that multiple tasks are trained simultaneously on a dataset~\cite{Caruana1997,articleq,9392366}. In other words, multiple cost functions are required to be minimized at the same time. Let us denote the cost function for the $t$-th task is $L_t(\theta)$ where $\theta$ is the parameter shared by all tasks~(for our purpose we only consider the case that all parameters are shared). It is important to note that those tasks may conflict and $\{L_t(\theta)\}$ cannot be minimized to their minimum values simultaneously. In other words, there should be some trade-off between those tasks. One well-accepted solution for  multi-objective optimization is to achieve Pareto optimality at optimized $\theta^*$, which can be stated as follows,
\begin{enumerate}
	\item $\theta$ dominates $\theta'$ if $L_t(\theta)\le L_t(\theta')$ for all $t$. 
	\item $\theta^*$ is Pareto optimal if there is no other $\theta$ dominates $\theta^*$.
\end{enumerate}
The Pareto optimality thus respects welfare for all tasks without sacrificing one.

A necessary condition for achieving the Pareto optimal solution is the Karush-Kuhn-Tucker~(KKT) conditions~\cite{Kuhn2014}: there exists $\{\alpha_t\}$ such that
\begin{equation}
    \label{eq:KKT}
    \sum_{t=1}^T\alpha_t=1,~~~G(\theta, \alpha)\equiv\sum_{t=1}^T\alpha_t\nabla_\theta L_t(\theta)=0, 
\end{equation}
where $\alpha_t \ge 0$ is the weight of the $t$-th task. Any solution that satisfies KKT conditions is called a Pareto stationary point.  It can be seen that the linear combination of gradients $G(\theta, \alpha)$ plays a central role for multi-objective optimization.

For achieving the Pareto optimal point,  we can use the multiple  gradient descent algorithm(MGDA)~\cite{NEURIPS2018_432aca3a,DESIDERI2012313}. At the $k$-th iteration, both the weights $\{\alpha_t\}$ and the parameters $\theta$ should be updated as following:
\begin{enumerate}
	\item  Finding the optimal weight  $\{\alpha_t\}$ for each task by solving the optimization problem,
	\begin{equation} 
\label{eq:sum_gradients_square}
\begin{aligned}
    &\min_{\alpha_1,\dots,\alpha_T}||\sum_{t=1}^T\alpha_t\nabla_\theta L_t(\theta)||^2\\
    &\text{s.t.} \sum_{t=1}^T\alpha_t=1, \alpha_t\geq 0.
\end{aligned}
\end{equation}
Finding the optimal weights is a quadratic optimization problem in a convex hull, which can be solved efficiently with a Frank-Wolfe algorithm~\cite{pmlr-v28-jaggi13}.
	\item  If $ G(\theta, \alpha)\neq0$, then uses a gradient descent to update $\theta$ as, 
	\begin{equation}
	\theta\leftarrow\theta-\eta \sum_{t=1}^T\alpha_t\nabla_\theta L_t(\theta)
	\end{equation}	
where $\eta$ is the step size. 
\end{enumerate}

At the step of determining the optimal weight  $\{\alpha_t\}$, minimization of a linear combination of gradients has the following consequence. For one specific task, if the norm of the gradient is large(small) the cost function is multiplied by a small(large) weight. This gives rise to a balance among those tasks in optimization. 

\begin{figure}[h]
\centering
\includegraphics[width=1.0\linewidth]{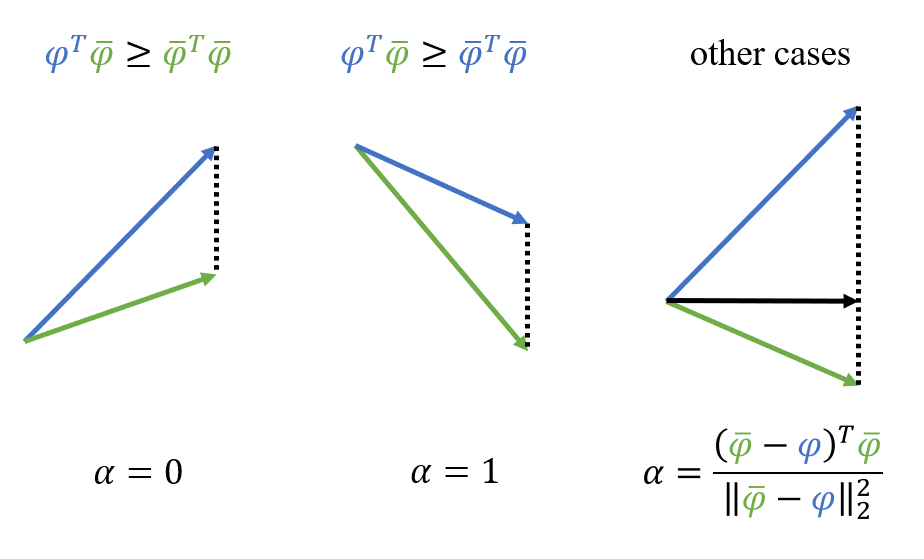}
\captionsetup{justification=raggedright, singlelinecheck=false}
\caption{Visualisation of the solution of (\ref{1}). Here $\varphi_i=\nabla_\theta L_i(\theta)$ for $i=1,2$.}
\label{fig:alpha}
\end{figure}

In this work, we only need to consider two tasks~($T=2$). One is for minimizing energy or free energy, and the other enforces the Gauss law.  In this case, determining the optimal weights can be solved exactly. The optimization problem in Eq.~\eqref{eq:sum_gradients_square}can be rewritten as 
\begin{equation}
    \label{1}
    \min_{\alpha\in[0,1]}||\alpha\nabla_\theta L_1(\theta)+(1-\alpha)\nabla_\theta L_2(\theta)||^2_2
\end{equation}
The solution of $\alpha$ is illustrated and given in Fig.~\ref{fig:alpha}. Finally, we use gradient descent method and update the parameters:
\begin{equation}
    \label{}
    \theta \leftarrow \theta-\eta[\alpha\nabla_\theta L_1(\theta)+(1-\alpha)\nabla_\theta L_2(\theta)]
\end{equation}

We now discuss how the method of multiple gradient descent algorithm for multi-objective optimization can be readily incorporated into variational quantum algorithms. VQA refers to a hybrid quantum-classical algorithm for optimization, where observables related to cost function and gradient are measured on quantum computer while parameters are updated by classical computing.  Thus, in multiple gradient descent algorithm,  evaluations of  each cost function $ L_t(\theta)$ and the corresponding gradient $\nabla_\theta L_t(\theta)$ are executed by quantum computing while determining the weights and update $\theta$ are  conducted classically. 
In the next section, we showcase how multi-objective optimization is utilized for preparing ground states and thermal states of lattice gauge theory with variational quantum algorithms. 

\section{\label{sec:two} Variational quantum algorithm for lattice gauge field }
In this section, we first introduce a one-dimensional lattice gauge theory, where spinless fermions are coupled to $Z_2$ gauge theory. Then we propose variational quantum eigensolver and variational quantum thermalizer for preparing the ground state and the thermal states for the one dimensional $Z_2$ LGT, respectively, by treating minimizing the energy or free energy and the enforcement of Gauss law as two objectives for optimization.

\subsection{One dimensional $Z_2$ lattice gauge theory}
We consider a chain where spinless fermions live on sites and $Z_2$ gauge field is defined on links~\cite{Borla_2020} .The quantum Hamiltonian of the system is,
\begin{equation}
    \label{1}
    H=-t\sum_j(c^\dagger_j\sigma_{j,j+1}^xc_{j+1}+h.c.)-h\sum_j\sigma_{j,j+1}^z
\end{equation}
where $c_j$ is fermionic operator and the Pauli operator $\sigma$ denotes $Z_2$ gauge field. Moreover, $\sigma_{j,j+1}^z$ is the electric field, while magnetic field does not exist in one dimension.  Notably, the fermionic matter and the gauge field should satisfy the Gauss law at each spatial site,
\begin{equation}
    G_j=\sigma_{j-1,j}^z(-1)^{c_j^\dagger c_j}\sigma_{j,j+1}^z=1. 
\end{equation}
The physical meaning of Gauss law is that the electric field changes only when there resides a charge (fermion). 

For digital quantum simulation of the one-dimensional $Z_2$ lattice gauge theory, it is necessary to map both the fermionic matter and the gauge field into qubits.  By Jordan-Wigner transformation, fermions can be mapped into qubits as $c_j={\textstyle\prod_{l=1}^{j-1}}(i\sigma_{l}^{z})\sigma_{j}^{-}$, where $\sigma^{\pm}=\frac{1}{2}(\sigma^{x}\pm i\sigma^{y})$. The $Z_2$ gauge field on one link can be directly represented by a qubit. For convenience, we denote qubit for fermions on the odd-number lattice while qubit for the gauge field on the even-number lattice. 

Then the qubit Hamiltonian can be written as follows, 
\begin{equation}
    H=-t\sum_j(\sigma^+_{2j-1}\sigma_{2j}^x\sigma^-_{2j+1}+h.c.)-h\sum_i\sigma_{2j}^z.
\end{equation}
The Gauss law in the qubit language can be rewritten as, 
\begin{equation}
    G_{j}=\sigma_{2j}^z\sigma_{2j+1}^z\sigma_{2j+2}^z=1.
\end{equation}
The Gauss law implies that all physical states should be eigenstates with $+1$ eigenvalues for all $G_{j}$ simultaneously. 
This gives a strong constraint on the physical Hilbert space. In fact, preserving the gauge invariance is one of the major challenges in quantum simulating lattice gauge theories on quantum hardware.

\begin{figure}[h]
\centering
\includegraphics[width=1.0\linewidth]{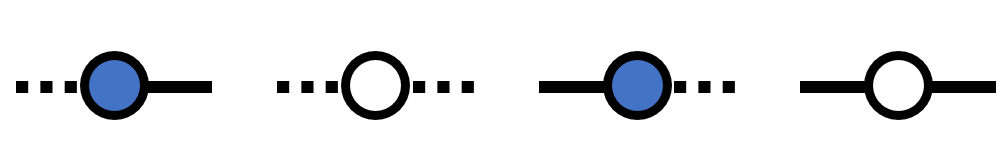}
\captionsetup{justification=raggedright, singlelinecheck=false}
\caption{The four basic configurations that satisfy the Gauss law, dashed (solid) lines mean $\sigma_z=1$($-1$) and filled~(empty) circles mean $n_i\equiv c_j^\dagger c_j=1$($0$). }
\label{Gauss}
\end{figure}

We use a spatial site with two links to illustrate the consequence of Gauss law on the allowed computational basis. Constrained by the Gauss law, there are four configurations, as shown  in FIG. \ref{Gauss}.

It is clear to see that the number of computational basis respecting the Gauss law will be much less than the dimension of Hilbert space. Nevertheless, the dimension of physical Hilbert space is still exponentially growing with the system size. In fact,  for a system with $N$ sites, the dimension of physical Hilbert space is $2^{N+1}$ under periodic condition.   For verifying the results of variational quantum algorithms, we can use exact diagonalization for calculating the ground state as well as the thermal state, by choosing all computational basis in the physical Hilbert space. The exact diagonalization for LGT is done using the package \emph{QuSpin}\cite{10.21468/SciPostPhys.2.1.003,10.21468/SciPostPhys.7.2.020}, where computational basis respecting the given constraints can be generated systematically.

\subsection{Variational quantum eigensolver}
\begin{figure}[h]
\centering
\includegraphics[width=0.5\linewidth]{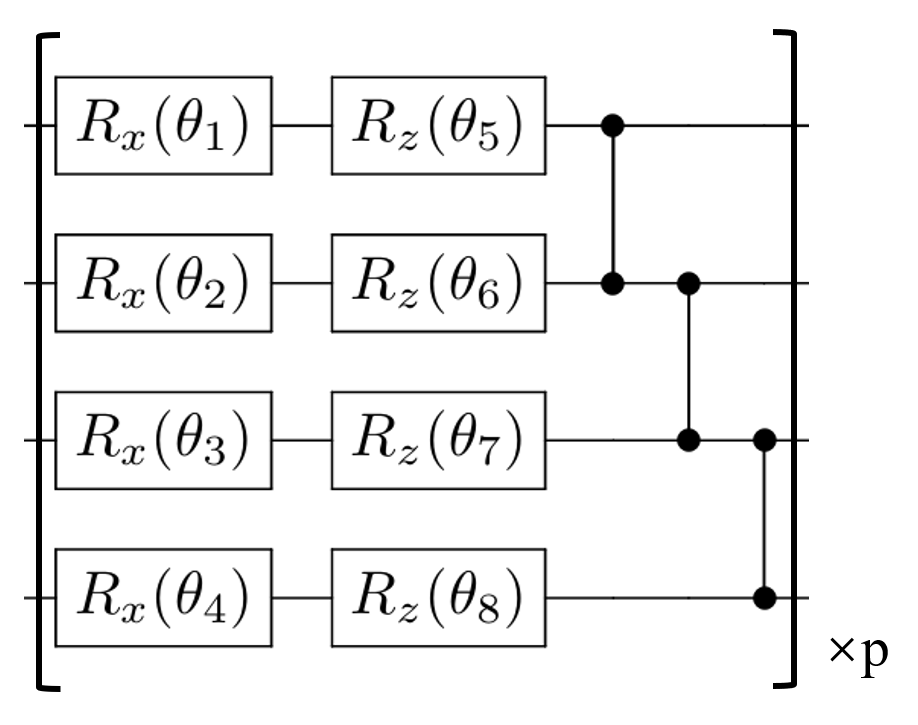}
\captionsetup{justification=raggedright, singlelinecheck=false}
\caption{A depiction of the universal ansatz for a system of four qubits.}
\label{ansatz}
\end{figure}

We first focus on the ground state preparation with variational quantum eigensolver.  With a parameterized unitary operator $U(\theta)$, the trial ground state can be expressed as 
\begin{equation}
    \label{}
    |\psi_\theta\rangle=U(\theta)|0\rangle^{\otimes n}
\end{equation}
Here we use a universal ansatz with two layers of one-qubit gates and one layer of two-qubit gates in each block, which is represented in FIG. \ref{ansatz}. The variational energy can be written as $E(\theta)=\langle\psi_\theta|H|\psi_\theta\rangle$. For typical quantum systems, one can obtain the ground state by minimizing $E(\theta)$.

For lattice gauge theory, it is not enough by just minimizing the variational energy,  as the goal is to obtain the lowest-energy eigenstate which also satisfies the Gauss law, e.g., the ground state of $H$ is also an eigenstate of $G_i$ with an eigenvalue $+1$.  We can consider it as a constrained optimization problem. One approach for constrained optimization is to use the penalty function method~\cite{doi:10.1137/0327068}. Utilizing penalty for variational quantum eigensolver can be found in Ref.~\cite{PhysRevResearch.3.013197,article5}. For lattice gauge theory, we can set a penalty for enforcing Gauss law at each site as$ |\langle\psi_\theta|G_i|\psi_\theta\rangle-1|$,  which is larger than zero when the trial state $|\psi_\theta\rangle$ violating the Gauss law.  By adding penalties for violating Gauss law at all sites, the total cost function becomes,
\begin{equation}
    \label{3}
    \mathcal{L}(\theta)= \langle\psi_\theta|H|\psi_\theta\rangle + \mu\sum_i^N|\langle\psi_\theta|G_i|\psi_\theta\rangle-1|.
\end{equation}
Here $\mu>0$ is the strength of the penalty. Note that choosing a proper $\mu$ is important but not easy. When $\mu$ is too small, optimization may lead to a state with energy lower than the true ground state by violating the Gauss law. Another approach, which is the focus of this paper, is to take minimization of variational energy and satisfying the Gauss law as two-objective optimization, whose cost function can be written as,
\begin{eqnarray}
    \mathcal{L}_1(\theta)&=& \langle\psi_\theta|H|\psi_\theta\rangle, \\ ~~ \mathcal{L}_2(\theta)&=&\sum_i^N|\langle\psi_\theta|G_i|\psi_\theta\rangle-1|.
\end{eqnarray}
Then we can use the multiple gradient descent algorithm for minimizing both $\mathcal{L}_1(\theta)$ and $\mathcal{L}_2(\theta)$. The detailed algorithm is given in Algorithm.~\ref{alg1}. A Pareto optimal solution can be expected that guarantees that energy is minimized under the constraint $\mathcal{L}_2(\theta)=0$. 

\begin{algorithm} \SetKwData{Left}{left}\SetKwData{This}{this}\SetKwData{Up}{up} \SetKwFunction{Union}{Union}\SetKwFunction{FindCompress}{FindCompress} \SetKwInOut{Input}{input}\SetKwInOut{Output}{output}
  \caption{VQE}
  \label{alg1}
    \Input{Parameter $\theta$ to be optimized, Hamiltonian $H$ and initial state $|\psi_0\rangle$.} 
    \Output{Optimization parameters $\theta^*$.}
    \BlankLine
    Construct a quantum circuit $U(\theta)$;

    Apply $U(\theta)$ to the initial state $|\psi_0\rangle$, then obtain $|\psi_\theta\rangle=U(\theta)|\psi_0\rangle$;

    $\alpha=\arg\min||\alpha\nabla_\theta \mathcal{L}_1(\theta)+(1-\alpha)\nabla_\theta \mathcal{L}_2(\theta)||^2_2$;

    $\theta \leftarrow \theta-\eta[\alpha\nabla_\theta \mathcal{L}_1(\theta)+(1-\alpha)\nabla_\theta \mathcal{L}_2(\theta)]$;

    Repeat step 3 and 4 until convergence.
\end{algorithm}

\subsection{Variational quantum thermalizer}

Now we extend the algorithm for lattice gauge theory to finite temperature. The goal is to obtain the thermal state. First of all, we discuss how exact thermal state as well as other thermodynamic observables can be obtained by exact diagonalization. By introducing a projecting operator $P$ into the physical Hilbert space constraint by the Gauss law, the Hamiltonian written in the physical Hilbert space should be expressed as $\Tilde{H}=PHP$. Then the thermal state at temperature $T$ can be written as, 
\begin{equation}
    \label{}
    \rho = \frac{e^{-\beta \Tilde{H}}}{\text{Tr}(e^{-\beta \Tilde{H}})},
\end{equation}
where $\beta=T^{-1}$ is the inverse temperature.  The exact free energy can be obtained as, 
\begin{equation}
    \label{}
    F=\text{Tr}(\rho\Tilde{H})-T S
\end{equation}
where $S=-\text{Tr}(\rho\log\rho)$ is the entropy. 

For variational preparing thermal state on quantum processors, one difficulty is to evaluate the variational free energy which involves measuring the entropy $S$. Unlike the energy which can often be obtained by measuring local observable, $S$ is a statistical observable and replies on the whole information of $\rho$. Here we adopt a product ansatz where the entropy is determined with an initial density matrix $\rho_\phi$ and can be calculated analytically~\cite{Xie_2022}. The initial density matrix  can be written as an uncorrelated tensor product of $n$ one-qubit density matrices $ \rho(\phi)=\otimes_i^n\rho_i(\phi_i)$, where
\begin{equation}
    \label{}
    \rho_i(\phi_i)=\sin^{2}\phi_{i}|0\rangle\langle0|+\cos^{2}\phi_{i}|1\rangle\langle1|
\end{equation}
Such one-qubit mixed state can be obtained by discarding one qubit of a two-qubit entangled state, $\cos\phi_{i}\left | 00 \right \rangle+\sin \phi_{i}\left | 11 \right \rangle$, which can be readily prepared on quantum processors. 

We consider the parameterized quantum circuit $U(\theta)$ with the structure illustrated in FIG. \ref{ansatz}. The trial thermal state with respect to parameters $\phi$ and $\theta$ can be written as~\cite{verdon2019quantumhamiltonianbasedmodelsvariational},
\begin{equation}
    \label{}
    \rho(\phi,\theta)=U(\theta)\rho(\phi) U^\dagger(\theta).
\end{equation}
Because the unitary transformation  $U(\theta)$ will not change the entropy, the entropy depends only on the initial state $\rho(\phi)$, which can be calculated as, 
\begin{equation}
    S(\phi)=\sum_i-\sin^{2}\phi_{i}\log\sin^{2}\phi_{i}-\cos^{2}\phi_{i} \log\cos^{2}\phi_{i}.
\end{equation}

For conventional quantum systems, it suffices to minimize the variational free energy for obtaining the thermal state.   However,  one should consider the Gauss law, by restricting the trial thermal state in the physical Hilbert space. One candidate approach is to add a penalty term to punish the violation of Gauss law. The cost function can be expressed as
\begin{equation}
    \label{}
    \mathcal{L}(\phi,\theta)=\text{Tr}(\rho(\phi,\theta)H)-TS(\phi)+\mu\sum_{i=1}^N|\text{Tr}\rho(\phi,\theta)G_i-1|.
\end{equation}
However, by simulating as shown later, we find that it works only when the temperature is extremely low. When the temperature rises to an intermediate temperature, such as $\beta=1$, this algorithm doesn't work. The reason is that the penalty coefficient $\mu$ is difficult to determine at non-zero temperature, as there is a competition among the energy, entropy and the penalty.  If the penalty is too small, then it tends to increase the entropy by allowing states in sectors $G_i=-1$ which violates the Gauss law. For too large $\mu$, the Gauss law may be satisfied at priority, while the accuracy of free energy will be lowered. To solve the problem, we will focus on the multi-objective optimization, where the two cost functions can be written respectively as, 
\begin{eqnarray}
   \mathcal{L}_1(\phi,\theta)&=&\text{Tr}(\rho(\phi,\theta)H)-TS(\phi) \\ ~~   \mathcal{L}_2(\phi,\theta)&=&\sum_{i=1}^N|\text{Tr}(\rho(\phi,\theta)G_i-1|.
\end{eqnarray}
The optimization of $\mathcal{L}_1(\phi,\theta)$ and $\mathcal{L}_2(\phi,\theta)$ using multiple gradient descent algorithm is similar to that of VQE, with a goal to obtaining the thermal state respecting the Gauss law. The detailed algorithm is given in Algorithm.~\ref{alg2}.

\begin{algorithm} \SetKwData{Left}{left}\SetKwData{This}{this}\SetKwData{Up}{up} \SetKwFunction{Union}{Union}\SetKwFunction{FindCompress}{FindCompress} \SetKwInOut{Input}{input}\SetKwInOut{Output}{output}
  \caption{VQT}
  \label{alg2}
    \Input{Parameter $\phi,\theta$ to be optimized, Hamiltonian $H$ and initial state $\rho_\phi$.} 
    \Output{Optimization parameters $\phi^*,\theta^*$.}
    \BlankLine
    Construct a quantum circuit $U(\theta)$;

    Apply $U(\theta)$ to the initial state $\rho_\phi$, then obtain $\rho(\phi,\theta)=U(\theta)\rho_\phi U^\dagger(\theta)$;

    $\alpha=\arg\min(||\alpha\nabla_{\phi} \mathcal{L}_1(\phi,\theta)+(1-\alpha)\nabla_{\phi} \mathcal{L}_2(\phi,\theta)||^2_2+||\alpha\nabla_{\theta} \mathcal{L}_1(\phi,\theta)+(1-\alpha)\nabla_{\theta} \mathcal{L}_2(\phi,\theta)||^2_2)$;

    $\phi \leftarrow \phi-\eta[\alpha\nabla_\phi \mathcal{L}_1(\phi,\theta)+(1-\alpha)\nabla_\phi \mathcal{L}_2(\phi,\theta)]$;

    $\theta \leftarrow \theta-\eta[\alpha\nabla_\theta \mathcal{L}_1(\phi,\theta)+(1-\alpha)\nabla_\theta \mathcal{L}_2(\phi,\theta)]$;

    Repeat step 3, 4 and 5 until convergence.
\end{algorithm}

\section{\label{sec:three}  Simulation Results}
In this section, we present simulation results for ground state and thermal state of a 1D lattice gauge model.

\subsection{\label{sec:oneCond}Ground State}
\begin{figure}[h]
\centering
\includegraphics[width=1.0\linewidth]{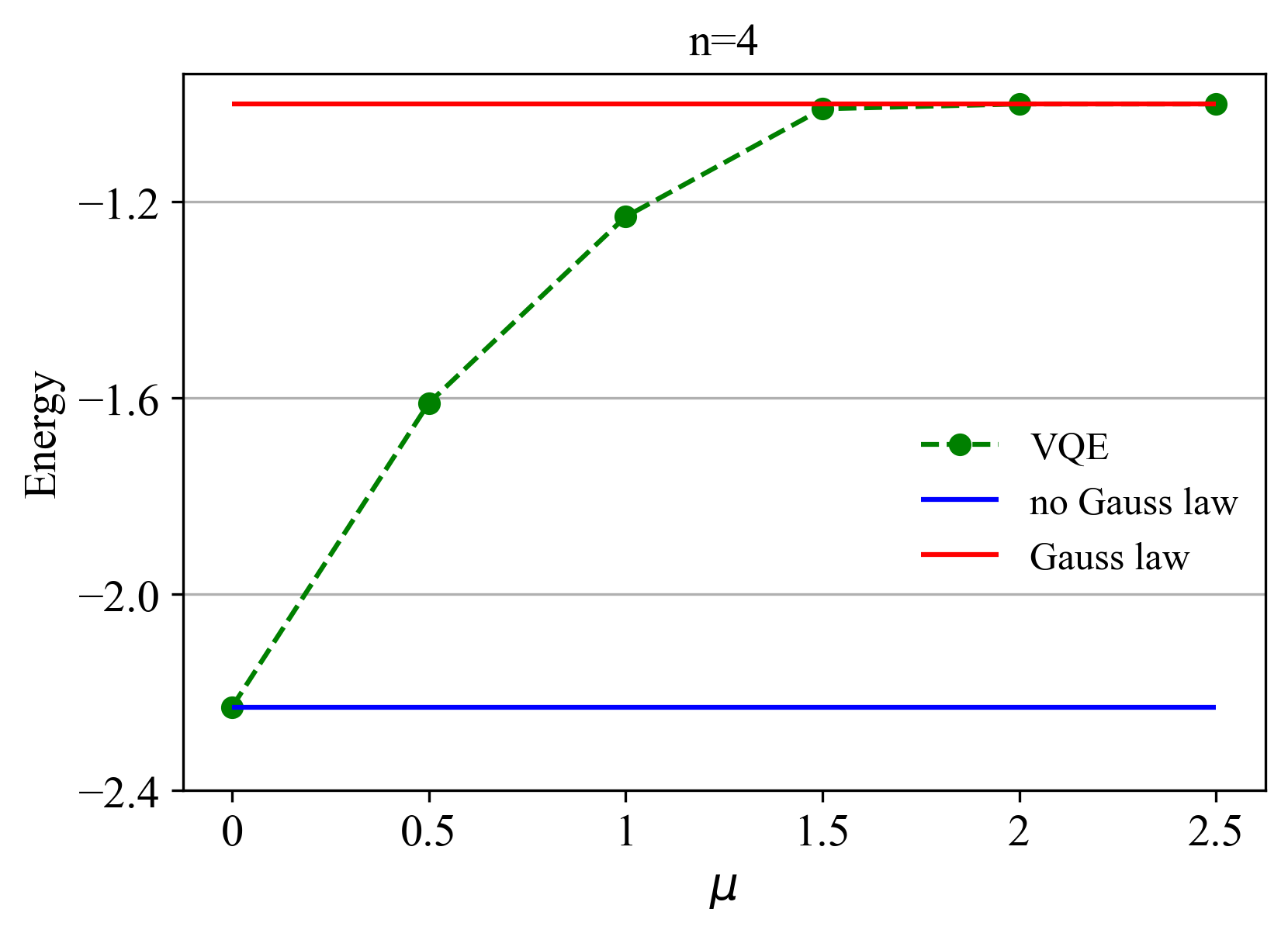}
\includegraphics[width=1.0\linewidth]{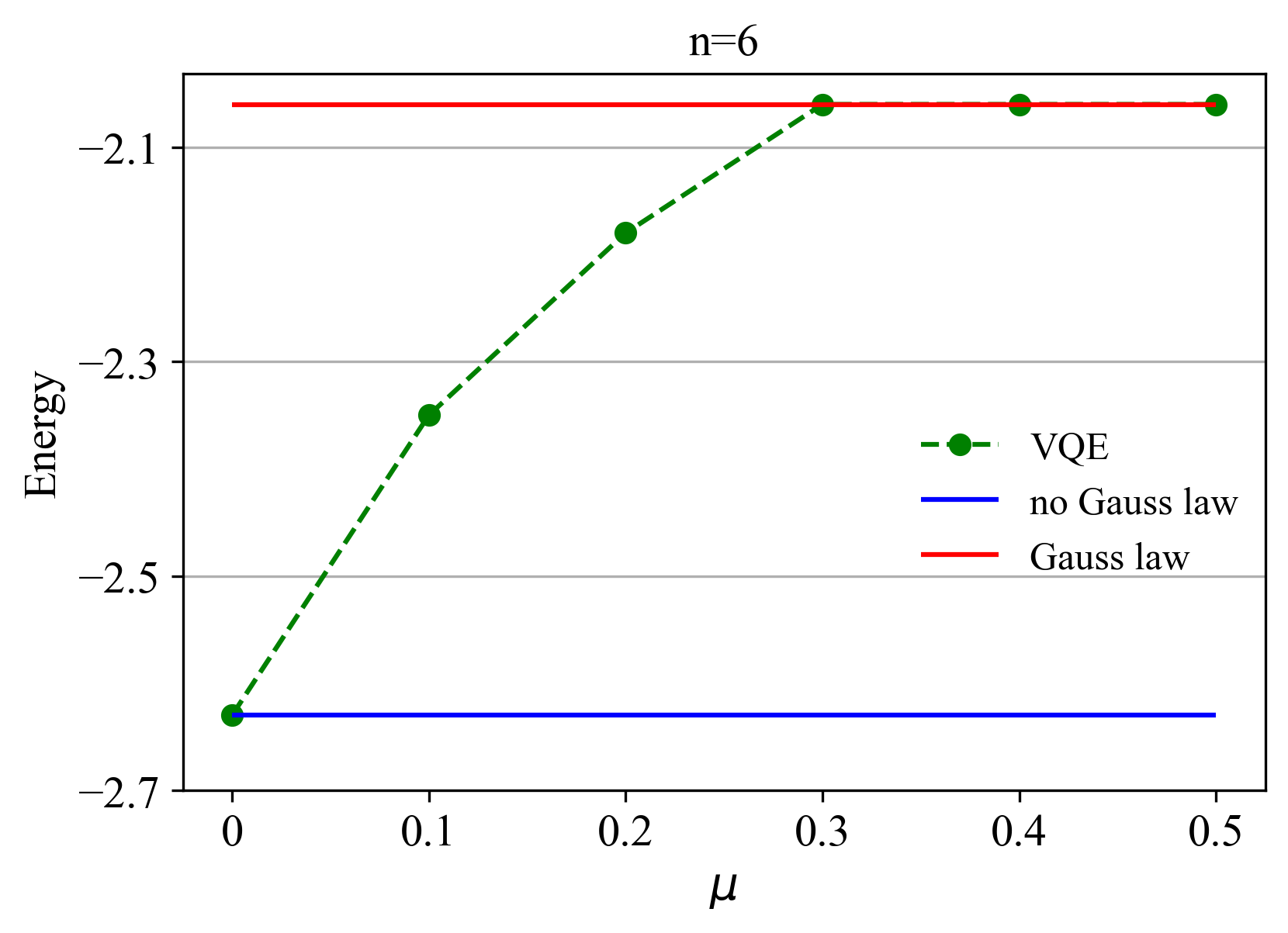}
\captionsetup{justification=raggedright, singlelinecheck=false}
\caption{The energy expectation value for different penalty coefficients, showing that as the penalty coefficient $\mu$ increases, the energy will converge to the ground-state energy that satisfies the Gauss law.}
\label{penalty}
\end{figure}

First, we focus on the ground state. Here, we consider two cases, 2 sites and 3 sites. Because there's a link between sites, there are 4 and 6 qubits, respectively. We first use VQE to obtain the ground-state energy which doesn't consider the Gauss law, and then we calculate the ground-state energy that considers the Gauss law as the exact value of the algorithm. The cost function of the algorithm is presented in (\ref{3}). We set different penalty coefficients and then minimize the cost function. The algorithm results are presented in FIG. \ref{penalty}. Here, the number of blocks of ansatz is $p=3$ and the step size is $\eta=0.02$ . We set $t=1$, $h=0.5$ in  the Hamiltonian. The optimization is done using the package \emph{scipy}~\cite{2020SciPy-NMeth}, where the cost function can be minimized.

By adding penalty terms in the cost function and magnifying penalty coefficient $\mu$, the energy will increase from the ground-state energy with no Gauss law to the ground-state energy with Gauss law. However, we find that how to determine $\mu$ is a problem, as analyzed in Section \ref{sec:two} and there exist a range of values. There is some discussion in \cite{PhysRevResearch.3.013197}.

\begin{figure}[h]
\centering
\includegraphics[width=1.0\linewidth]{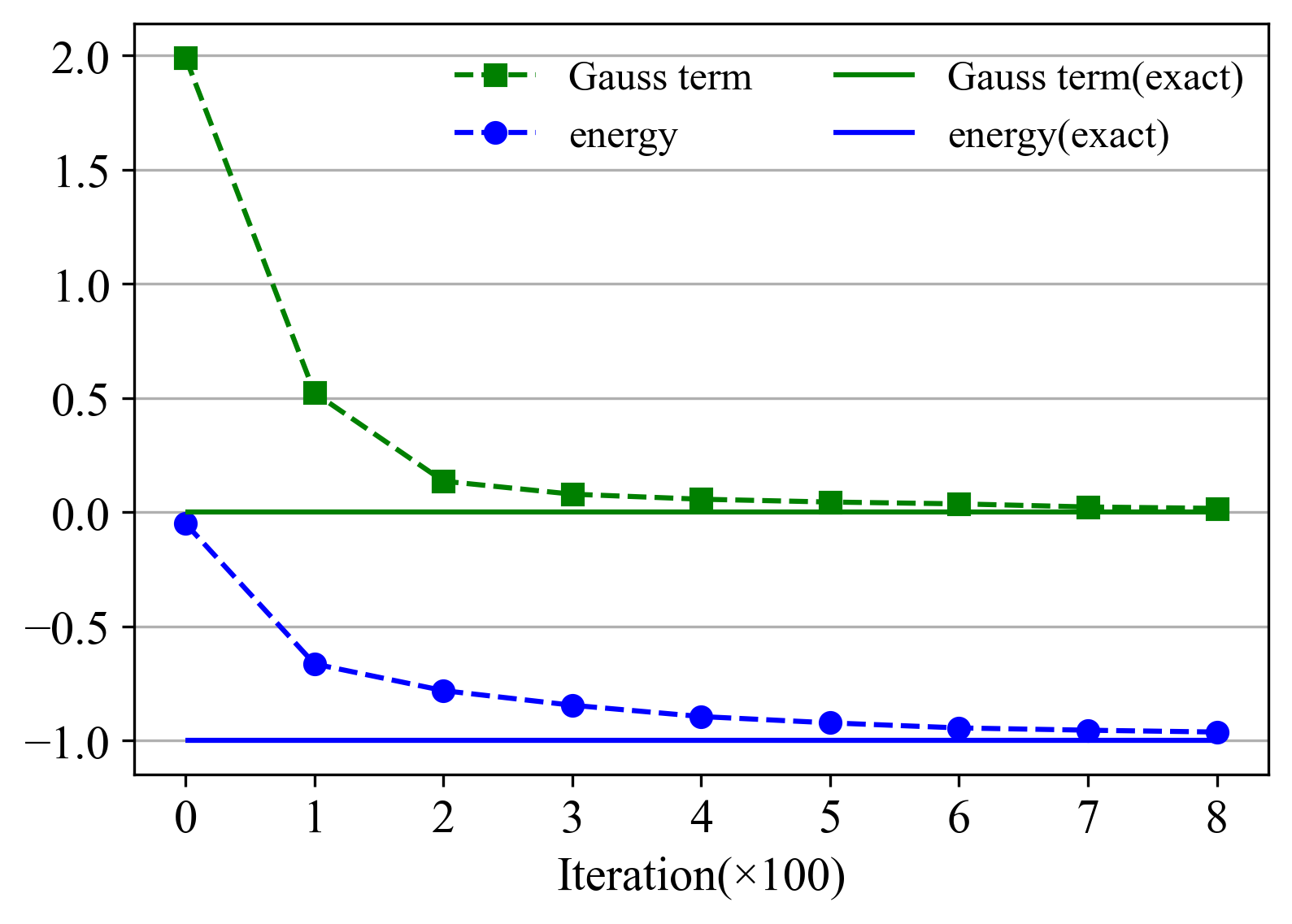}
\captionsetup{justification=raggedright, singlelinecheck=false}
\caption{Multi-objective optimizaiton for variational quantum eigensolver of 1D lattice gauge theory. The exact value of Gauss law is 0 which means that the state satisfies the Gauss law. The exact value of energy is the minimum that the state should satisfies the Gauss law.}
\label{ground}
\end{figure}

Because it's hard to determine the penalty coefficient and our goal is to minimize the energy expectation value, and meanwhile, the quantum state must also satisfies the Gauss law. Therefore, we consider this problem as a multi-objective problem. It's clear that the minimum of the Gauss term is 0 and when it is 0, the eigenvalue of $G_i$ will be 1. It means that the state has satisfied the Gauss law constraints. We use multiple gradient descent algorithm to optimize two cost functions simultaneously. The results are presented in FIG. \ref{ground}.

\begin{figure}[h]
\centering
\includegraphics[width=1.0\linewidth]{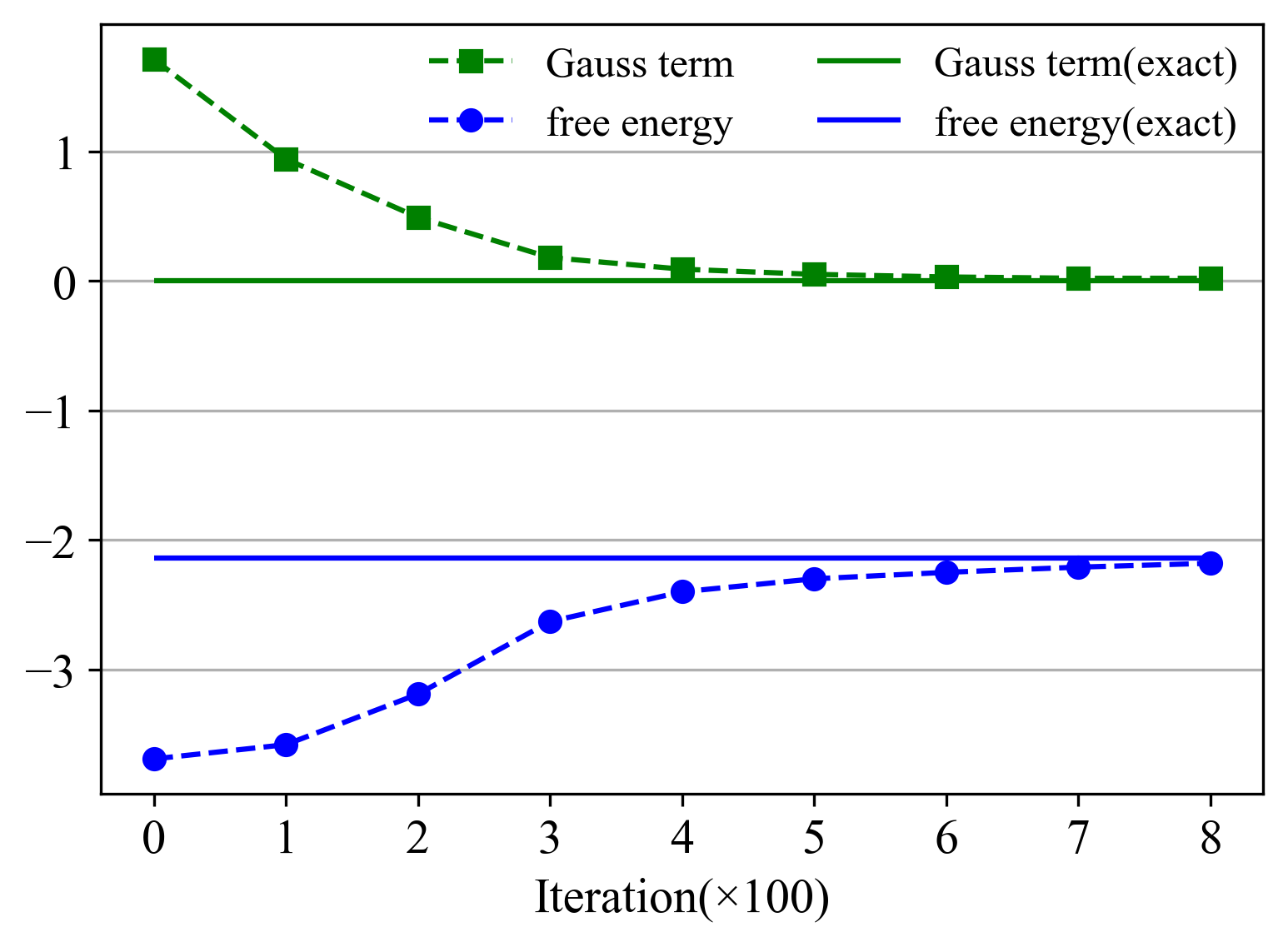}
\captionsetup{justification=raggedright, singlelinecheck=false}
\caption{Multi-objective optimization for variational quantum thermalizer of 1D lattice gauge theory. The temperature is set as $T=1$. The exact value of Gauss law is 0 which means that the state satisfies the Gauss law. The exact value of free energy is the minimum that the state satisfies the Gauss law.}
\label{thermal 1}
\end{figure}

It's clear that as the number of iterations increases, $\mathcal{L}_2(\theta)$ will converge to 0, which means that the state we obtained from algorithm has satisfied the Gauss law. Also, the energy expectation value will decrease and finally also converge to its exact value.

\subsection{\label{sec:oneCond}Thermal State}
Now we turn to variational preparing thermal state of the 1D $Z_2$ lattice gauge theory at non-zero temperature.  The goal is to minimize the free energy, and meanwhile, the quantum state must also satisfies the Gauss law. Here, the number of blocks of ansatz is $p=3$. We set a small step size  $\eta=0.02$ and the initial parameters are chosen randomly.   We present the gradient descent result in FIG. \ref{thermal 1}.

\begin{figure}[h]
\centering
\includegraphics[width=1\linewidth]{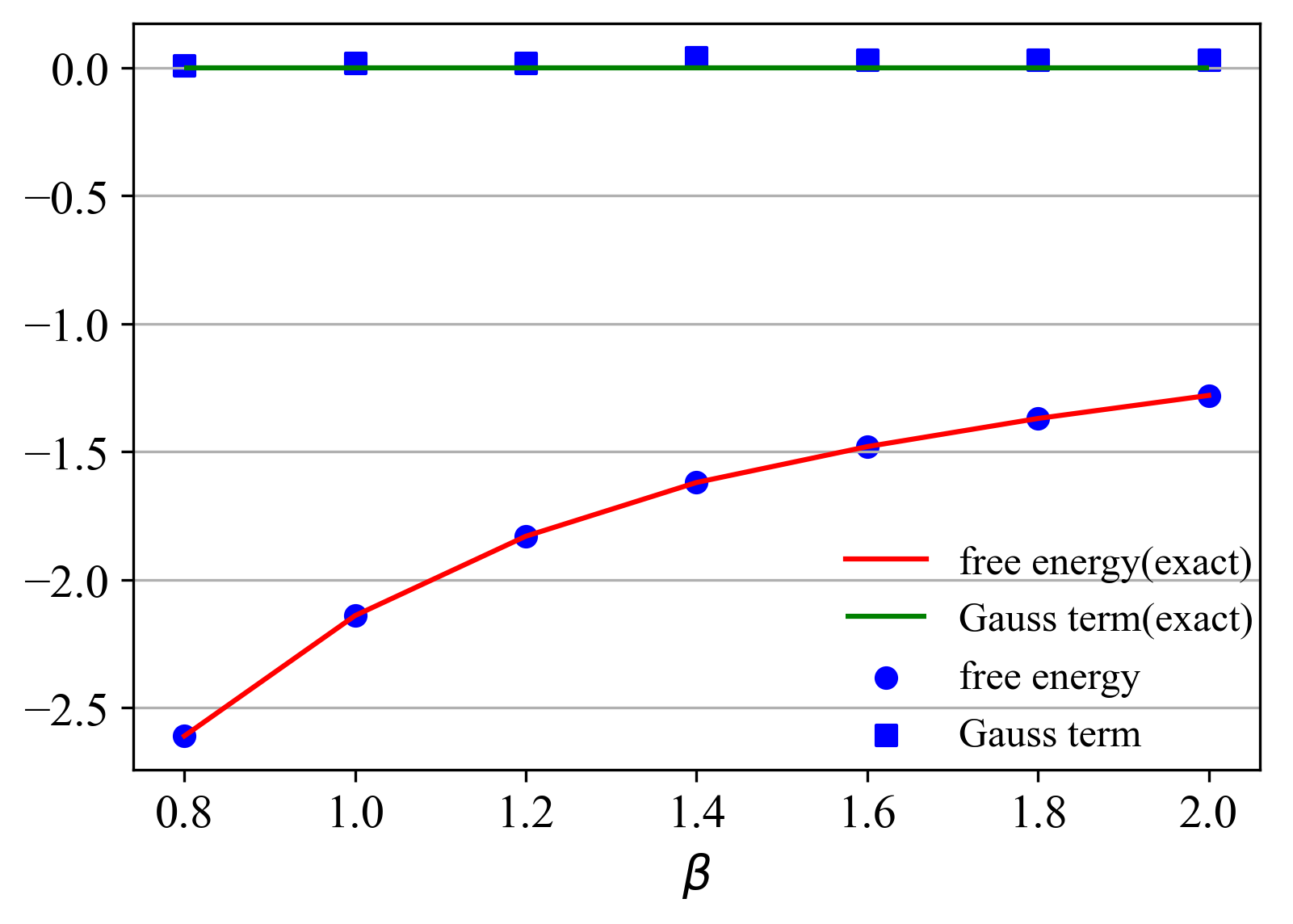}
\captionsetup{justification=raggedright, singlelinecheck=false}
\caption{Free energy at different temperatures obtained by variational quantum thermalizer with multi-objective optimization. The solid lines are the exact value and dots are the results obtained by VQT.}
\label{thermal 2}
\end{figure}

\begin{figure}[h]
\centering
\subfigure[]{
    \includegraphics[width=0.48\linewidth]{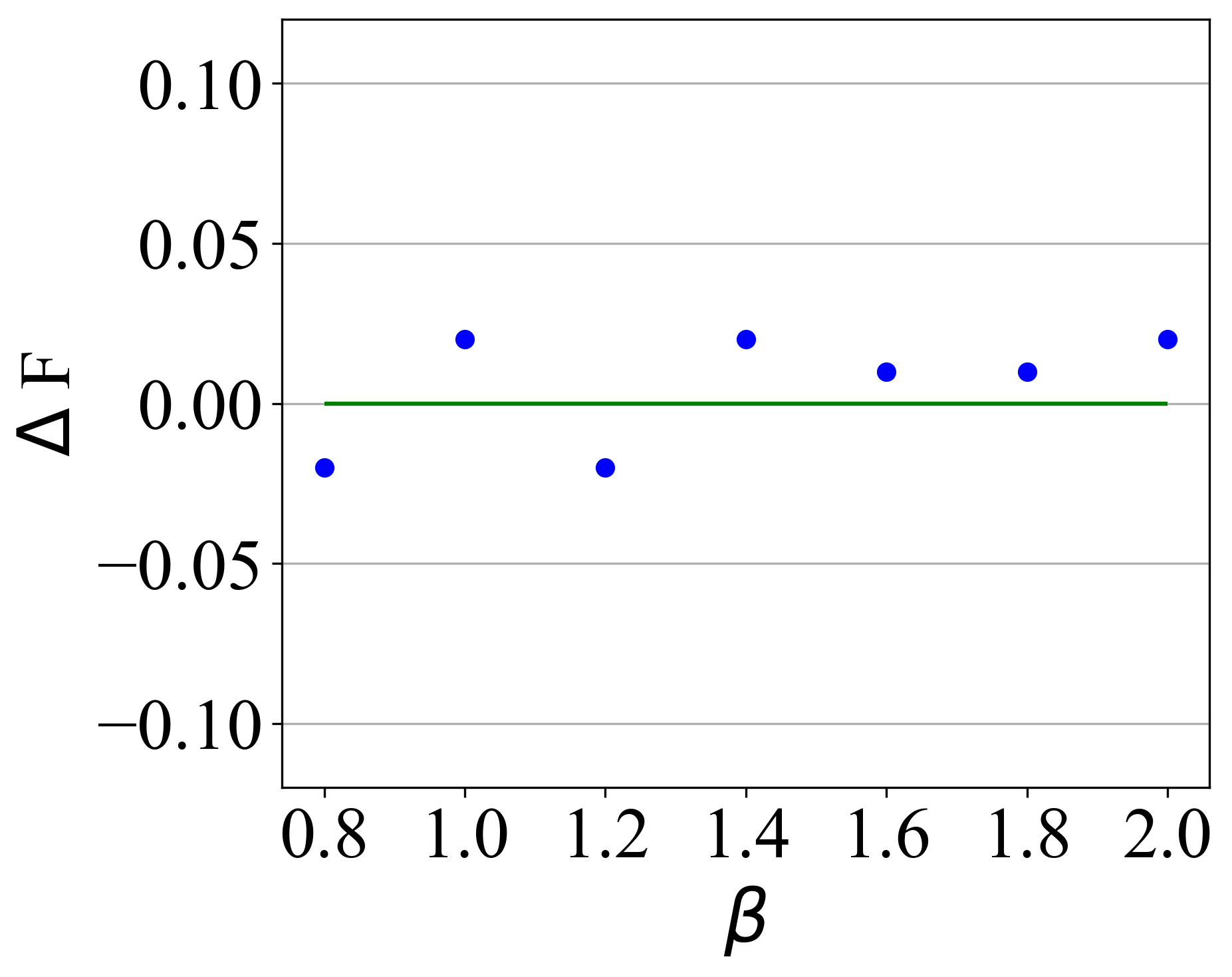}}
\subfigure[]{
    \includegraphics[width=0.48\linewidth]{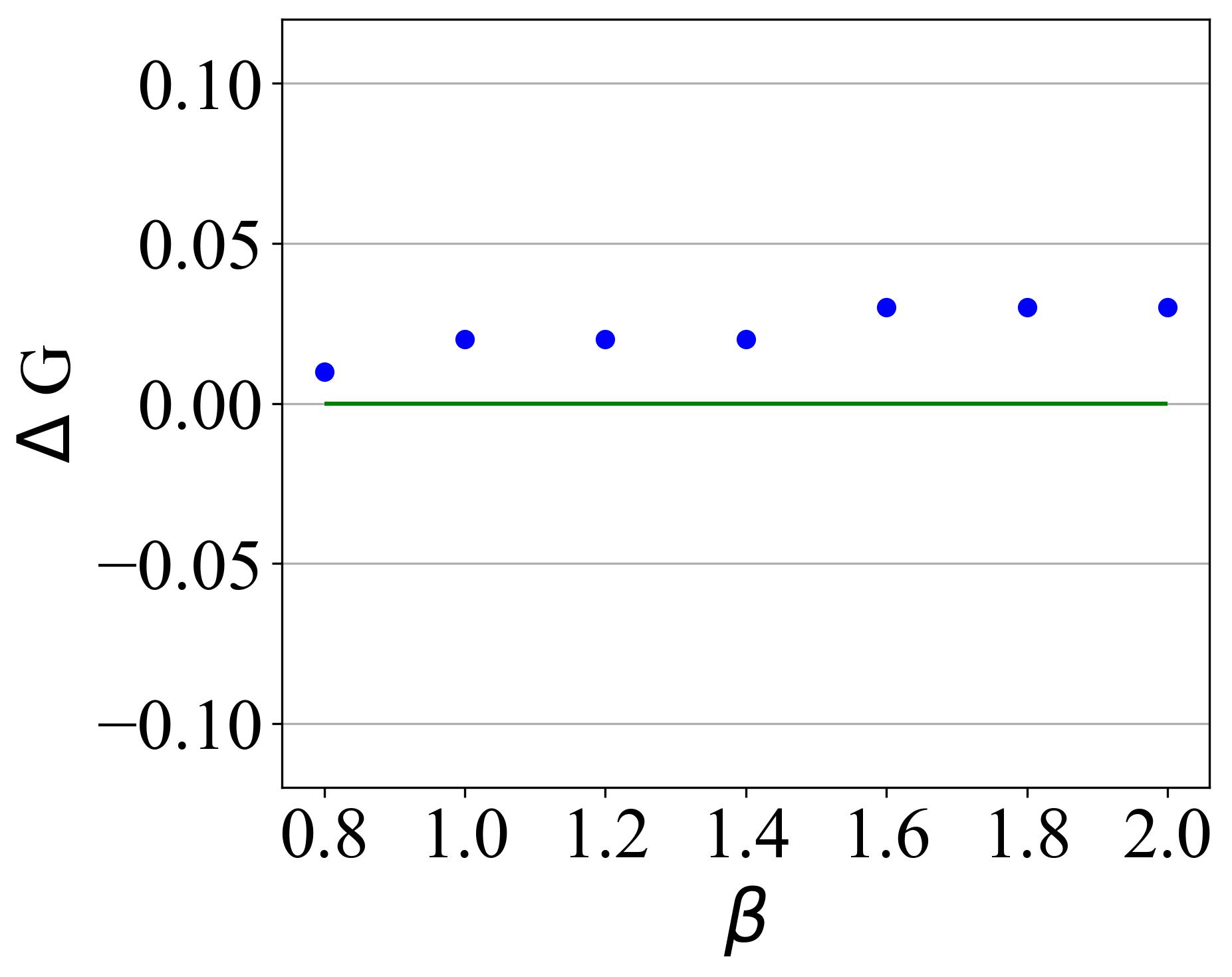}}
\captionsetup{justification=raggedright, singlelinecheck=false}
\caption{Error of variational quantum thermalizer for lattice gauge theory. (a) The difference between exact value and free energy obtained by VQT. (b)The difference between exact value and Gauss terms.}
\label{thermal 3}
\end{figure}

It is clear that as the number of iterations increases, $\mathcal{L}_2(\phi,\theta)$ will converge to 0, which means that the thermal state we obtained from algorithm has satisfied the Gauss law. Also, the free energy  increases and finally also converges to its exact value. The results show that as gradient descent continues, the free energy will start increasing from a small value and eventually converge to the exact value we calculated in advance, which seems inconsistent because the goal of gradient descent is to reduce a function. The reason is that we are minimizing two functions simultaneously here, which share a set of parameters. As $\mathcal{L}_2(\phi,\theta)$ gradually approaches 0, it means that the state increasingly satisfies the Gauss law, and the Hilbert space is approaching to physical Hilbert space. The free energy of the thermal state will be greater than that of the thermal state that does not satisfy the Gauss law. Therefore, as the number of iteration increases, the free energy will increase rather than decrease. 

Besides fixing a temperature, we also simulate the 1D LGT at different temperatures. Remarkably, we focus on some intermediate temperature, where the penalty function methods fail to work. The simulation result is presented in FIG. \ref{thermal 2} and the free energy is very close to its exact value. Also, they all satisfy the Gauss law because every $\mathcal{L}_2(\phi,\theta)$ is close to 0. We also show the difference between the exact value and free energy or Gauss terms in FIG. \ref{thermal 3}

\section{\label{sec:four}Conclusions}
In summary, we have proposed an approach for solving lattice gauge field problems that combines variational quantum algorithm and multi-objective optimization . We have given a procedure for reaching the Pareto optimal for preparing the ground state and thermal state constraint by the Gauss law by multiple gradient descent algorithm. Demonstrated with a 1D $Z_2$ lattice gauge theory, we have shown with numeral simulation that the algorithm can accurately prepare both the ground state and thermal states. Our work provides a variational quantum approach for simulating the lattice gauge theory by quantum computing.

\begin{acknowledgments}
%%%%%%%%%%%%%%%%%%%%%%%
 This work was supported by the National Natural Science Foundation of China (Grant No.12375013) and the Guangdong Basic and Applied Basic Research Fund (Grant No.2023A1515011460).
\end{acknowledgments}

\normalem
\balance
\bibliography{LangxingCheng}

\begin{thebibliography}{48}%
\makeatletter
\providecommand \@ifxundefined [1]{%
 \@ifx{#1\undefined}
}%
\providecommand \@ifnum [1]{%
 \ifnum #1\expandafter \@firstoftwo
 \else \expandafter \@secondoftwo
 \fi
}%
\providecommand \@ifx [1]{%
 \ifx #1\expandafter \@firstoftwo
 \else \expandafter \@secondoftwo
 \fi
}%
\providecommand \natexlab [1]{#1}%
\providecommand \enquote  [1]{``#1''}%
\providecommand \bibnamefont  [1]{#1}%
\providecommand \bibfnamefont [1]{#1}%
\providecommand \citenamefont [1]{#1}%
\providecommand \href@noop [0]{\@secondoftwo}%
\providecommand \href [0]{\begingroup \@sanitize@url \@href}%
\providecommand \@href[1]{\@@startlink{#1}\@@href}%
\providecommand \@@href[1]{\endgroup#1\@@endlink}%
\providecommand \@sanitize@url [0]{\catcode `\\12\catcode `\$12\catcode `\&12\catcode `\#12\catcode `\^12\catcode `\_12\catcode `\%12\relax}%
\providecommand \@@startlink[1]{}%
\providecommand \@@endlink[0]{}%
\providecommand \url  [0]{\begingroup\@sanitize@url \@url }%
\providecommand \@url [1]{\endgroup\@href {#1}{\urlprefix }}%
\providecommand \urlprefix  [0]{URL }%
\providecommand \Eprint [0]{\href }%
\providecommand \doibase [0]{https://doi.org/}%
\providecommand \selectlanguage [0]{\@gobble}%
\providecommand \bibinfo  [0]{\@secondoftwo}%
\providecommand \bibfield  [0]{\@secondoftwo}%
\providecommand \translation [1]{[#1]}%
\providecommand \BibitemOpen [0]{}%
\providecommand \bibitemStop [0]{}%
\providecommand \bibitemNoStop [0]{.\EOS\space}%
\providecommand \EOS [0]{\spacefactor3000\relax}%
\providecommand \BibitemShut  [1]{\csname bibitem#1\endcsname}%
\let\auto@bib@innerbib\@empty
%</preamble>
\bibitem [{\citenamefont {Motta}\ \emph {et~al.}(2019)\citenamefont {Motta}, \citenamefont {Sun}, \citenamefont {Tan}, \citenamefont {O’Rourke}, \citenamefont {Ye}, \citenamefont {Minnich}, \citenamefont {Brandão},\ and\ \citenamefont {Chan}}]{Motta_2019}%
  \BibitemOpen
  \bibfield  {author} {\bibinfo {author} {\bibfnamefont {M.}~\bibnamefont {Motta}}, \bibinfo {author} {\bibfnamefont {C.}~\bibnamefont {Sun}}, \bibinfo {author} {\bibfnamefont {A.~T.~K.}\ \bibnamefont {Tan}}, \bibinfo {author} {\bibfnamefont {M.~J.}\ \bibnamefont {O’Rourke}}, \bibinfo {author} {\bibfnamefont {E.}~\bibnamefont {Ye}}, \bibinfo {author} {\bibfnamefont {A.~J.}\ \bibnamefont {Minnich}}, \bibinfo {author} {\bibfnamefont {F.~G. S.~L.}\ \bibnamefont {Brandão}},\ and\ \bibinfo {author} {\bibfnamefont {G.~K.-L.}\ \bibnamefont {Chan}},\ }\bibfield  {title} {\bibinfo {title} {Determining eigenstates and thermal states on a quantum computer using quantum imaginary time evolution},\ }\href {https://doi.org/10.1038/s41567-019-0704-4} {\bibfield  {journal} {\bibinfo  {journal} {Nature Physics}\ }\textbf {\bibinfo {volume} {16}},\ \bibinfo {pages} {205–210} (\bibinfo {year} {2019})}\BibitemShut {NoStop}%
\bibitem [{\citenamefont {McArdle}\ \emph {et~al.}(2019)\citenamefont {McArdle}, \citenamefont {Jones}, \citenamefont {Endo}, \citenamefont {Li}, \citenamefont {Benjamin},\ and\ \citenamefont {Yuan}}]{McArdle_2019}%
  \BibitemOpen
  \bibfield  {author} {\bibinfo {author} {\bibfnamefont {S.}~\bibnamefont {McArdle}}, \bibinfo {author} {\bibfnamefont {T.}~\bibnamefont {Jones}}, \bibinfo {author} {\bibfnamefont {S.}~\bibnamefont {Endo}}, \bibinfo {author} {\bibfnamefont {Y.}~\bibnamefont {Li}}, \bibinfo {author} {\bibfnamefont {S.~C.}\ \bibnamefont {Benjamin}},\ and\ \bibinfo {author} {\bibfnamefont {X.}~\bibnamefont {Yuan}},\ }\bibfield  {title} {\bibinfo {title} {Variational ansatz-based quantum simulation of imaginary time evolution},\ }\bibfield  {journal} {\bibinfo  {journal} {npj Quantum Information}\ }\textbf {\bibinfo {volume} {5}},\ \href {https://doi.org/10.1038/s41534-019-0187-2} {10.1038/s41534-019-0187-2} (\bibinfo {year} {2019})\BibitemShut {NoStop}%
\bibitem [{\citenamefont {Ren}\ \emph {et~al.}(2022)\citenamefont {Ren}, \citenamefont {Fu}, \citenamefont {Wu},\ and\ \citenamefont {Chen}}]{Ren2022TowardsTG}%
  \BibitemOpen
  \bibfield  {author} {\bibinfo {author} {\bibfnamefont {W.}~\bibnamefont {Ren}}, \bibinfo {author} {\bibfnamefont {W.}~\bibnamefont {Fu}}, \bibinfo {author} {\bibfnamefont {X.}~\bibnamefont {Wu}},\ and\ \bibinfo {author} {\bibfnamefont {J.}~\bibnamefont {Chen}},\ }\bibfield  {title} {\bibinfo {title} {Towards the ground state of molecules via diffusion monte carlo on neural networks},\ }\href {https://api.semanticscholar.org/CorpusID:248476061} {\bibfield  {journal} {\bibinfo  {journal} {Nature Communications}\ }\textbf {\bibinfo {volume} {14}} (\bibinfo {year} {2022})}\BibitemShut {NoStop}%
\bibitem [{\citenamefont {Cerezo}\ \emph {et~al.}(2021)\citenamefont {Cerezo}, \citenamefont {Arrasmith}, \citenamefont {Babbush}, \citenamefont {Benjamin}, \citenamefont {Endo}, \citenamefont {Fujii}, \citenamefont {McClean}, \citenamefont {Mitarai}, \citenamefont {Yuan}, \citenamefont {Cincio},\ and\ \citenamefont {Coles}}]{Cerezo_2021}%
  \BibitemOpen
  \bibfield  {author} {\bibinfo {author} {\bibfnamefont {M.}~\bibnamefont {Cerezo}}, \bibinfo {author} {\bibfnamefont {A.}~\bibnamefont {Arrasmith}}, \bibinfo {author} {\bibfnamefont {R.}~\bibnamefont {Babbush}}, \bibinfo {author} {\bibfnamefont {S.~C.}\ \bibnamefont {Benjamin}}, \bibinfo {author} {\bibfnamefont {S.}~\bibnamefont {Endo}}, \bibinfo {author} {\bibfnamefont {K.}~\bibnamefont {Fujii}}, \bibinfo {author} {\bibfnamefont {J.~R.}\ \bibnamefont {McClean}}, \bibinfo {author} {\bibfnamefont {K.}~\bibnamefont {Mitarai}}, \bibinfo {author} {\bibfnamefont {X.}~\bibnamefont {Yuan}}, \bibinfo {author} {\bibfnamefont {L.}~\bibnamefont {Cincio}},\ and\ \bibinfo {author} {\bibfnamefont {P.~J.}\ \bibnamefont {Coles}},\ }\bibfield  {title} {\bibinfo {title} {Variational quantum algorithms},\ }\href {https://doi.org/10.1038/s42254-021-00348-9} {\bibfield  {journal} {\bibinfo  {journal} {Nature Reviews Physics}\ }\textbf {\bibinfo {volume} {3}},\ \bibinfo {pages} {625–644} (\bibinfo {year} {2021})}\BibitemShut
  {NoStop}%
\bibitem [{\citenamefont {Huang}\ \emph {et~al.}(2023)\citenamefont {Huang}, \citenamefont {Xu}, \citenamefont {Guo}, \citenamefont {Tian}, \citenamefont {Wei}, \citenamefont {Sun}, \citenamefont {Bao},\ and\ \citenamefont {Long}}]{Huang_2023}%
  \BibitemOpen
  \bibfield  {author} {\bibinfo {author} {\bibfnamefont {H.-L.}\ \bibnamefont {Huang}}, \bibinfo {author} {\bibfnamefont {X.-Y.}\ \bibnamefont {Xu}}, \bibinfo {author} {\bibfnamefont {C.}~\bibnamefont {Guo}}, \bibinfo {author} {\bibfnamefont {G.}~\bibnamefont {Tian}}, \bibinfo {author} {\bibfnamefont {S.-J.}\ \bibnamefont {Wei}}, \bibinfo {author} {\bibfnamefont {X.}~\bibnamefont {Sun}}, \bibinfo {author} {\bibfnamefont {W.-S.}\ \bibnamefont {Bao}},\ and\ \bibinfo {author} {\bibfnamefont {G.-L.}\ \bibnamefont {Long}},\ }\bibfield  {title} {\bibinfo {title} {Near-term quantum computing techniques: Variational quantum algorithms, error mitigation, circuit compilation, benchmarking and classical simulation},\ }\bibfield  {journal} {\bibinfo  {journal} {Science China Physics, Mechanics and Astronomy}\ }\textbf {\bibinfo {volume} {66}},\ \href {https://doi.org/10.1007/s11433-022-2057-y} {10.1007/s11433-022-2057-y} (\bibinfo {year} {2023})\BibitemShut {NoStop}%
\bibitem [{\citenamefont {Qi}\ \emph {et~al.}(2024)\citenamefont {Qi}, \citenamefont {Xiao}, \citenamefont {Liu}, \citenamefont {Gong},\ and\ \citenamefont {Gani}}]{article}%
  \BibitemOpen
  \bibfield  {author} {\bibinfo {author} {\bibfnamefont {H.}~\bibnamefont {Qi}}, \bibinfo {author} {\bibfnamefont {S.}~\bibnamefont {Xiao}}, \bibinfo {author} {\bibfnamefont {Z.}~\bibnamefont {Liu}}, \bibinfo {author} {\bibfnamefont {C.}~\bibnamefont {Gong}},\ and\ \bibinfo {author} {\bibfnamefont {A.}~\bibnamefont {Gani}},\ }\bibfield  {title} {\bibinfo {title} {Variational quantum algorithms: fundamental concepts, applications and challenges},\ }\href {https://doi.org/10.1007/s11128-024-04438-2} {\bibfield  {journal} {\bibinfo  {journal} {Quantum Information Processing}\ }\textbf {\bibinfo {volume} {23}} (\bibinfo {year} {2024})}\BibitemShut {NoStop}%
\bibitem [{\citenamefont {Preskill}(2018)}]{Preskill_2018}%
  \BibitemOpen
  \bibfield  {author} {\bibinfo {author} {\bibfnamefont {J.}~\bibnamefont {Preskill}},\ }\bibfield  {title} {\bibinfo {title} {Quantum computing in the nisq era and beyond},\ }\href {https://doi.org/10.22331/q-2018-08-06-79} {\bibfield  {journal} {\bibinfo  {journal} {Quantum}\ }\textbf {\bibinfo {volume} {2}},\ \bibinfo {pages} {79} (\bibinfo {year} {2018})}\BibitemShut {NoStop}%
\bibitem [{\citenamefont {Bharti}\ \emph {et~al.}(2022)\citenamefont {Bharti}, \citenamefont {Cervera-Lierta}, \citenamefont {Kyaw}, \citenamefont {Haug}, \citenamefont {Alperin-Lea}, \citenamefont {Anand}, \citenamefont {Degroote}, \citenamefont {Heimonen}, \citenamefont {Kottmann}, \citenamefont {Menke}, \citenamefont {Mok}, \citenamefont {Sim}, \citenamefont {Kwek},\ and\ \citenamefont {Aspuru-Guzik}}]{Bharti_2022}%
  \BibitemOpen
  \bibfield  {author} {\bibinfo {author} {\bibfnamefont {K.}~\bibnamefont {Bharti}}, \bibinfo {author} {\bibfnamefont {A.}~\bibnamefont {Cervera-Lierta}}, \bibinfo {author} {\bibfnamefont {T.~H.}\ \bibnamefont {Kyaw}}, \bibinfo {author} {\bibfnamefont {T.}~\bibnamefont {Haug}}, \bibinfo {author} {\bibfnamefont {S.}~\bibnamefont {Alperin-Lea}}, \bibinfo {author} {\bibfnamefont {A.}~\bibnamefont {Anand}}, \bibinfo {author} {\bibfnamefont {M.}~\bibnamefont {Degroote}}, \bibinfo {author} {\bibfnamefont {H.}~\bibnamefont {Heimonen}}, \bibinfo {author} {\bibfnamefont {J.~S.}\ \bibnamefont {Kottmann}}, \bibinfo {author} {\bibfnamefont {T.}~\bibnamefont {Menke}}, \bibinfo {author} {\bibfnamefont {W.-K.}\ \bibnamefont {Mok}}, \bibinfo {author} {\bibfnamefont {S.}~\bibnamefont {Sim}}, \bibinfo {author} {\bibfnamefont {L.-C.}\ \bibnamefont {Kwek}},\ and\ \bibinfo {author} {\bibfnamefont {A.}~\bibnamefont {Aspuru-Guzik}},\ }\bibfield  {title} {\bibinfo {title} {Noisy intermediate-scale quantum algorithms},\ }\bibfield
  {journal} {\bibinfo  {journal} {Reviews of Modern Physics}\ }\textbf {\bibinfo {volume} {94}},\ \href {https://doi.org/10.1103/revmodphys.94.015004} {10.1103/revmodphys.94.015004} (\bibinfo {year} {2022})\BibitemShut {NoStop}%
\bibitem [{\citenamefont {Benedetti}\ \emph {et~al.}(2019)\citenamefont {Benedetti}, \citenamefont {Lloyd}, \citenamefont {Sack},\ and\ \citenamefont {Fiorentini}}]{Benedetti_2019}%
  \BibitemOpen
  \bibfield  {author} {\bibinfo {author} {\bibfnamefont {M.}~\bibnamefont {Benedetti}}, \bibinfo {author} {\bibfnamefont {E.}~\bibnamefont {Lloyd}}, \bibinfo {author} {\bibfnamefont {S.}~\bibnamefont {Sack}},\ and\ \bibinfo {author} {\bibfnamefont {M.}~\bibnamefont {Fiorentini}},\ }\bibfield  {title} {\bibinfo {title} {Parameterized quantum circuits as machine learning models},\ }\href {https://doi.org/10.1088/2058-9565/ab4eb5} {\bibfield  {journal} {\bibinfo  {journal} {Quantum Science and Technology}\ }\textbf {\bibinfo {volume} {4}},\ \bibinfo {pages} {043001} (\bibinfo {year} {2019})}\BibitemShut {NoStop}%
\bibitem [{\citenamefont {Du}\ \emph {et~al.}(2021)\citenamefont {Du}, \citenamefont {Hsieh}, \citenamefont {Liu}, \citenamefont {You},\ and\ \citenamefont {Tao}}]{PRXQuantum.2.040337}%
  \BibitemOpen
  \bibfield  {author} {\bibinfo {author} {\bibfnamefont {Y.}~\bibnamefont {Du}}, \bibinfo {author} {\bibfnamefont {M.-H.}\ \bibnamefont {Hsieh}}, \bibinfo {author} {\bibfnamefont {T.}~\bibnamefont {Liu}}, \bibinfo {author} {\bibfnamefont {S.}~\bibnamefont {You}},\ and\ \bibinfo {author} {\bibfnamefont {D.}~\bibnamefont {Tao}},\ }\bibfield  {title} {\bibinfo {title} {Learnability of quantum neural networks},\ }\href {https://doi.org/10.1103/PRXQuantum.2.040337} {\bibfield  {journal} {\bibinfo  {journal} {PRX Quantum}\ }\textbf {\bibinfo {volume} {2}},\ \bibinfo {pages} {040337} (\bibinfo {year} {2021})}\BibitemShut {NoStop}%
\bibitem [{\citenamefont {Abbas}\ \emph {et~al.}(2021)\citenamefont {Abbas}, \citenamefont {Sutter}, \citenamefont {Zoufal}, \citenamefont {Lucchi}, \citenamefont {Figalli},\ and\ \citenamefont {Woerner}}]{Abbas_2021}%
  \BibitemOpen
  \bibfield  {author} {\bibinfo {author} {\bibfnamefont {A.}~\bibnamefont {Abbas}}, \bibinfo {author} {\bibfnamefont {D.}~\bibnamefont {Sutter}}, \bibinfo {author} {\bibfnamefont {C.}~\bibnamefont {Zoufal}}, \bibinfo {author} {\bibfnamefont {A.}~\bibnamefont {Lucchi}}, \bibinfo {author} {\bibfnamefont {A.}~\bibnamefont {Figalli}},\ and\ \bibinfo {author} {\bibfnamefont {S.}~\bibnamefont {Woerner}},\ }\bibfield  {title} {\bibinfo {title} {The power of quantum neural networks},\ }\href {https://doi.org/10.1038/s43588-021-00084-1} {\bibfield  {journal} {\bibinfo  {journal} {Nature Computational Science}\ }\textbf {\bibinfo {volume} {1}},\ \bibinfo {pages} {403–409} (\bibinfo {year} {2021})}\BibitemShut {NoStop}%
\bibitem [{\citenamefont {Peruzzo}\ \emph {et~al.}(2014)\citenamefont {Peruzzo}, \citenamefont {Alberto}, \citenamefont {McClean}, \citenamefont {Jarrod}, \citenamefont {Shadbolt}, \citenamefont {Peter}, \citenamefont {Yung}, \citenamefont {Man-Hong}, \citenamefont {Zhou}, \citenamefont {Love}, \citenamefont {Aspuru-Guzik}, \citenamefont {Alán}, \citenamefont {O’Brien},\ and\ \citenamefont {L.}}]{Peruzzo2014}%
  \BibitemOpen
  \bibfield  {author} {\bibinfo {author} {\bibnamefont {Peruzzo}}, \bibinfo {author} {\bibnamefont {Alberto}}, \bibinfo {author} {\bibnamefont {McClean}}, \bibinfo {author} {\bibnamefont {Jarrod}}, \bibinfo {author} {\bibnamefont {Shadbolt}}, \bibinfo {author} {\bibnamefont {Peter}}, \bibinfo {author} {\bibnamefont {Yung}}, \bibinfo {author} {\bibnamefont {Man-Hong}}, \bibinfo {author} {\bibfnamefont {X.-Q.}\ \bibnamefont {Zhou}}, \bibinfo {author} {\bibfnamefont {P.~J.}\ \bibnamefont {Love}}, \bibinfo {author} {\bibnamefont {Aspuru-Guzik}}, \bibinfo {author} {\bibnamefont {Alán}}, \bibinfo {author} {\bibnamefont {O’Brien}},\ and\ \bibinfo {author} {\bibfnamefont {J.}~\bibnamefont {L.}},\ }\bibfield  {title} {\bibinfo {title} {A variational eigenvalue solver on a photonic quantum processor},\ }\bibfield  {journal} {\bibinfo  {journal} {Nature Communications}\ }\textbf {\bibinfo {volume} {5}},\ \href {https://doi.org/10.1038/ncomms5213} {10.1038/ncomms5213} (\bibinfo {year} {2014})\BibitemShut {NoStop}%
\bibitem [{\citenamefont {Fedorov}\ \emph {et~al.}(2021)\citenamefont {Fedorov}, \citenamefont {Peng}, \citenamefont {Govind},\ and\ \citenamefont {Alexeev}}]{Fedorov2021VQEMA}%
  \BibitemOpen
  \bibfield  {author} {\bibinfo {author} {\bibfnamefont {D.~A.}\ \bibnamefont {Fedorov}}, \bibinfo {author} {\bibfnamefont {B.}~\bibnamefont {Peng}}, \bibinfo {author} {\bibfnamefont {N.}~\bibnamefont {Govind}},\ and\ \bibinfo {author} {\bibfnamefont {Y.}~\bibnamefont {Alexeev}},\ }\bibfield  {title} {\bibinfo {title} {Vqe method: a short survey and recent developments},\ }\href {https://api.semanticscholar.org/CorpusID:232232828} {\bibfield  {journal} {\bibinfo  {journal} {Materials Theory}\ }\textbf {\bibinfo {volume} {6}},\ \bibinfo {pages} {1} (\bibinfo {year} {2021})}\BibitemShut {NoStop}%
\bibitem [{\citenamefont {Higgott}\ \emph {et~al.}(2019)\citenamefont {Higgott}, \citenamefont {Wang},\ and\ \citenamefont {Brierley}}]{Higgott2019variationalquantum}%
  \BibitemOpen
  \bibfield  {author} {\bibinfo {author} {\bibfnamefont {O.}~\bibnamefont {Higgott}}, \bibinfo {author} {\bibfnamefont {D.}~\bibnamefont {Wang}},\ and\ \bibinfo {author} {\bibfnamefont {S.}~\bibnamefont {Brierley}},\ }\bibfield  {title} {\bibinfo {title} {Variational {Q}uantum {C}omputation of {E}xcited {S}tates},\ }\href {https://doi.org/10.22331/q-2019-07-01-156} {\bibfield  {journal} {\bibinfo  {journal} {{Quantum}}\ }\textbf {\bibinfo {volume} {3}},\ \bibinfo {pages} {156} (\bibinfo {year} {2019})}\BibitemShut {NoStop}%
\bibitem [{\citenamefont {Nakanishi}\ \emph {et~al.}(2019)\citenamefont {Nakanishi}, \citenamefont {Mitarai},\ and\ \citenamefont {Fujii}}]{Nakanishi_2019}%
  \BibitemOpen
  \bibfield  {author} {\bibinfo {author} {\bibfnamefont {K.~M.}\ \bibnamefont {Nakanishi}}, \bibinfo {author} {\bibfnamefont {K.}~\bibnamefont {Mitarai}},\ and\ \bibinfo {author} {\bibfnamefont {K.}~\bibnamefont {Fujii}},\ }\bibfield  {title} {\bibinfo {title} {Subspace-search variational quantum eigensolver for excited states},\ }\bibfield  {journal} {\bibinfo  {journal} {Physical Review Research}\ }\textbf {\bibinfo {volume} {1}},\ \href {https://doi.org/10.1103/physrevresearch.1.033062} {10.1103/physrevresearch.1.033062} (\bibinfo {year} {2019})\BibitemShut {NoStop}%
\bibitem [{\citenamefont {Verdon}\ \emph {et~al.}(2019)\citenamefont {Verdon}, \citenamefont {Marks}, \citenamefont {Nanda}, \citenamefont {Leichenauer},\ and\ \citenamefont {Hidary}}]{verdon2019quantumhamiltonianbasedmodelsvariational}%
  \BibitemOpen
  \bibfield  {author} {\bibinfo {author} {\bibfnamefont {G.}~\bibnamefont {Verdon}}, \bibinfo {author} {\bibfnamefont {J.}~\bibnamefont {Marks}}, \bibinfo {author} {\bibfnamefont {S.}~\bibnamefont {Nanda}}, \bibinfo {author} {\bibfnamefont {S.}~\bibnamefont {Leichenauer}},\ and\ \bibinfo {author} {\bibfnamefont {J.}~\bibnamefont {Hidary}},\ }\href {https://arxiv.org/abs/1910.02071} {\bibinfo {title} {Quantum hamiltonian-based models and the variational quantum thermalizer algorithm}} (\bibinfo {year} {2019}),\ \Eprint {https://arxiv.org/abs/1910.02071} {arXiv:1910.02071 [quant-ph]} \BibitemShut {NoStop}%
\bibitem [{\citenamefont {Selisko}\ \emph {et~al.}(2023)\citenamefont {Selisko}, \citenamefont {Amsler}, \citenamefont {Hammerschmidt}, \citenamefont {Drautz},\ and\ \citenamefont {Eckl}}]{Selisko_2024}%
  \BibitemOpen
  \bibfield  {author} {\bibinfo {author} {\bibfnamefont {J.}~\bibnamefont {Selisko}}, \bibinfo {author} {\bibfnamefont {M.}~\bibnamefont {Amsler}}, \bibinfo {author} {\bibfnamefont {T.}~\bibnamefont {Hammerschmidt}}, \bibinfo {author} {\bibfnamefont {R.}~\bibnamefont {Drautz}},\ and\ \bibinfo {author} {\bibfnamefont {T.}~\bibnamefont {Eckl}},\ }\bibfield  {title} {\bibinfo {title} {Extending the variational quantum eigensolver to finite temperatures},\ }\href {https://doi.org/10.1088/2058-9565/ad1340} {\bibfield  {journal} {\bibinfo  {journal} {Quantum Science and Technology}\ }\textbf {\bibinfo {volume} {9}},\ \bibinfo {pages} {015026} (\bibinfo {year} {2023})}\BibitemShut {NoStop}%
\bibitem [{\citenamefont {Consiglio}\ \emph {et~al.}(2024)\citenamefont {Consiglio}, \citenamefont {Settino}, \citenamefont {Giordano}, \citenamefont {Mastroianni}, \citenamefont {Plastina}, \citenamefont {Lorenzo}, \citenamefont {Maniscalco}, \citenamefont {Goold},\ and\ \citenamefont {Apollaro}}]{Consiglio_2024}%
  \BibitemOpen
  \bibfield  {author} {\bibinfo {author} {\bibfnamefont {M.}~\bibnamefont {Consiglio}}, \bibinfo {author} {\bibfnamefont {J.}~\bibnamefont {Settino}}, \bibinfo {author} {\bibfnamefont {A.}~\bibnamefont {Giordano}}, \bibinfo {author} {\bibfnamefont {C.}~\bibnamefont {Mastroianni}}, \bibinfo {author} {\bibfnamefont {F.}~\bibnamefont {Plastina}}, \bibinfo {author} {\bibfnamefont {S.}~\bibnamefont {Lorenzo}}, \bibinfo {author} {\bibfnamefont {S.}~\bibnamefont {Maniscalco}}, \bibinfo {author} {\bibfnamefont {J.}~\bibnamefont {Goold}},\ and\ \bibinfo {author} {\bibfnamefont {T.~J.~G.}\ \bibnamefont {Apollaro}},\ }\bibfield  {title} {\bibinfo {title} {Variational gibbs state preparation on noisy intermediate-scale quantum devices},\ }\bibfield  {journal} {\bibinfo  {journal} {Physical Review A}\ }\textbf {\bibinfo {volume} {110}},\ \href {https://doi.org/10.1103/physreva.110.012445} {10.1103/physreva.110.012445} (\bibinfo {year} {2024})\BibitemShut {NoStop}%
\bibitem [{\citenamefont {Xie}\ \emph {et~al.}(2022)\citenamefont {Xie}, \citenamefont {Guo}, \citenamefont {Xing}, \citenamefont {Xue}, \citenamefont {Zhang},\ and\ \citenamefont {Zhu}}]{Xie_2022}%
  \BibitemOpen
  \bibfield  {author} {\bibinfo {author} {\bibfnamefont {X.-D.}\ \bibnamefont {Xie}}, \bibinfo {author} {\bibfnamefont {X.}~\bibnamefont {Guo}}, \bibinfo {author} {\bibfnamefont {H.}~\bibnamefont {Xing}}, \bibinfo {author} {\bibfnamefont {Z.-Y.}\ \bibnamefont {Xue}}, \bibinfo {author} {\bibfnamefont {D.-B.}\ \bibnamefont {Zhang}},\ and\ \bibinfo {author} {\bibfnamefont {S.-L.}\ \bibnamefont {Zhu}},\ }\bibfield  {title} {\bibinfo {title} {Variational thermal quantum simulation of the lattice schwinger model},\ }\bibfield  {journal} {\bibinfo  {journal} {Physical Review D}\ }\textbf {\bibinfo {volume} {106}},\ \href {https://doi.org/10.1103/physrevd.106.054509} {10.1103/physrevd.106.054509} (\bibinfo {year} {2022})\BibitemShut {NoStop}%
\bibitem [{\citenamefont {Ichinose}\ and\ \citenamefont {Matsui}(2014)}]{Ichinose_2014}%
  \BibitemOpen
  \bibfield  {author} {\bibinfo {author} {\bibfnamefont {I.}~\bibnamefont {Ichinose}}\ and\ \bibinfo {author} {\bibfnamefont {T.}~\bibnamefont {Matsui}},\ }\bibfield  {title} {\bibinfo {title} {Lattice gauge theory for condensed matter physics: ferromagnetic superconductivity as its example},\ }\href {https://doi.org/10.1142/s0217984914300129} {\bibfield  {journal} {\bibinfo  {journal} {Modern Physics Letters B}\ }\textbf {\bibinfo {volume} {28}},\ \bibinfo {pages} {1430012} (\bibinfo {year} {2014})}\BibitemShut {NoStop}%
\bibitem [{\citenamefont {Wilson}(1974)}]{PhysRevD.10.2445}%
  \BibitemOpen
  \bibfield  {author} {\bibinfo {author} {\bibfnamefont {K.~G.}\ \bibnamefont {Wilson}},\ }\bibfield  {title} {\bibinfo {title} {Confinement of quarks},\ }\href {https://doi.org/10.1103/PhysRevD.10.2445} {\bibfield  {journal} {\bibinfo  {journal} {Phys. Rev. D}\ }\textbf {\bibinfo {volume} {10}},\ \bibinfo {pages} {2445} (\bibinfo {year} {1974})}\BibitemShut {NoStop}%
\bibitem [{\citenamefont {Kogut}(1979)}]{RevModPhys.51.659}%
  \BibitemOpen
  \bibfield  {author} {\bibinfo {author} {\bibfnamefont {J.~B.}\ \bibnamefont {Kogut}},\ }\bibfield  {title} {\bibinfo {title} {An introduction to lattice gauge theory and spin systems},\ }\href {https://doi.org/10.1103/RevModPhys.51.659} {\bibfield  {journal} {\bibinfo  {journal} {Rev. Mod. Phys.}\ }\textbf {\bibinfo {volume} {51}},\ \bibinfo {pages} {659} (\bibinfo {year} {1979})}\BibitemShut {NoStop}%
\bibitem [{\citenamefont {Kogut}(1983)}]{RevModPhys.55.775}%
  \BibitemOpen
  \bibfield  {author} {\bibinfo {author} {\bibfnamefont {J.~B.}\ \bibnamefont {Kogut}},\ }\bibfield  {title} {\bibinfo {title} {The lattice gauge theory approach to quantum chromodynamics},\ }\href {https://doi.org/10.1103/RevModPhys.55.775} {\bibfield  {journal} {\bibinfo  {journal} {Rev. Mod. Phys.}\ }\textbf {\bibinfo {volume} {55}},\ \bibinfo {pages} {775} (\bibinfo {year} {1983})}\BibitemShut {NoStop}%
\bibitem [{\citenamefont {Yan}\ \emph {et~al.}(2023)\citenamefont {Yan}, \citenamefont {Zhou}, \citenamefont {Zhou}, \citenamefont {Wang}, \citenamefont {Qiu}, \citenamefont {Meng},\ and\ \citenamefont {Zhang}}]{Yan2023}%
  \BibitemOpen
  \bibfield  {author} {\bibinfo {author} {\bibfnamefont {Z.}~\bibnamefont {Yan}}, \bibinfo {author} {\bibfnamefont {Z.}~\bibnamefont {Zhou}}, \bibinfo {author} {\bibfnamefont {Y.-H.}\ \bibnamefont {Zhou}}, \bibinfo {author} {\bibfnamefont {Y.-C.}\ \bibnamefont {Wang}}, \bibinfo {author} {\bibfnamefont {X.}~\bibnamefont {Qiu}}, \bibinfo {author} {\bibfnamefont {Z.~Y.}\ \bibnamefont {Meng}},\ and\ \bibinfo {author} {\bibfnamefont {X.-F.~a.}\ \bibnamefont {Zhang}},\ }\bibfield  {title} {\bibinfo {title} {Quantum optimization within lattice gauge theory model on a quantum simulator},\ }\bibfield  {journal} {\bibinfo  {journal} {npj Quantum Information}\ }\textbf {\bibinfo {volume} {9}},\ \href {https://doi.org/10.1038/s41534-023-00755-z} {10.1038/s41534-023-00755-z} (\bibinfo {year} {2023})\BibitemShut {NoStop}%
\bibitem [{\citenamefont {Emonts}\ \emph {et~al.}(2020)\citenamefont {Emonts}, \citenamefont {Bañuls}, \citenamefont {Cirac},\ and\ \citenamefont {Zohar}}]{article2}%
  \BibitemOpen
  \bibfield  {author} {\bibinfo {author} {\bibfnamefont {P.}~\bibnamefont {Emonts}}, \bibinfo {author} {\bibfnamefont {M.-C.}\ \bibnamefont {Bañuls}}, \bibinfo {author} {\bibfnamefont {I.}~\bibnamefont {Cirac}},\ and\ \bibinfo {author} {\bibfnamefont {E.}~\bibnamefont {Zohar}},\ }\bibfield  {title} {\bibinfo {title} {Variational monte carlo simulation with tensor networks of a pure z 3 gauge theory in ( 2 + 1 ) d},\ }\href {https://doi.org/10.1103/PhysRevD.102.074501} {\bibfield  {journal} {\bibinfo  {journal} {Physical Review D}\ }\textbf {\bibinfo {volume} {102}} (\bibinfo {year} {2020})}\BibitemShut {NoStop}%
\bibitem [{\citenamefont {Emonts}\ \emph {et~al.}(2023)\citenamefont {Emonts}, \citenamefont {Kelman}, \citenamefont {Borla}, \citenamefont {Moroz}, \citenamefont {Gazit},\ and\ \citenamefont {Zohar}}]{PhysRevD.107.014505}%
  \BibitemOpen
  \bibfield  {author} {\bibinfo {author} {\bibfnamefont {P.}~\bibnamefont {Emonts}}, \bibinfo {author} {\bibfnamefont {A.}~\bibnamefont {Kelman}}, \bibinfo {author} {\bibfnamefont {U.}~\bibnamefont {Borla}}, \bibinfo {author} {\bibfnamefont {S.}~\bibnamefont {Moroz}}, \bibinfo {author} {\bibfnamefont {S.}~\bibnamefont {Gazit}},\ and\ \bibinfo {author} {\bibfnamefont {E.}~\bibnamefont {Zohar}},\ }\bibfield  {title} {\bibinfo {title} {Finding the ground state of a lattice gauge theory with fermionic tensor networks: A $2+1\mathrm{D}$ ${\mathbb{z}}_{2}$ demonstration},\ }\href {https://doi.org/10.1103/PhysRevD.107.014505} {\bibfield  {journal} {\bibinfo  {journal} {Phys. Rev. D}\ }\textbf {\bibinfo {volume} {107}},\ \bibinfo {pages} {014505} (\bibinfo {year} {2023})}\BibitemShut {NoStop}%
\bibitem [{\citenamefont {Zohar}\ and\ \citenamefont {Cirac}(2018)}]{article3}%
  \BibitemOpen
  \bibfield  {author} {\bibinfo {author} {\bibfnamefont {E.}~\bibnamefont {Zohar}}\ and\ \bibinfo {author} {\bibfnamefont {J.}~\bibnamefont {Cirac}},\ }\bibfield  {title} {\bibinfo {title} {Combining tensor networks with monte carlo methods for lattice gauge theories},\ }\href {https://doi.org/10.1103/PhysRevD.97.034510} {\bibfield  {journal} {\bibinfo  {journal} {Physical Review D}\ }\textbf {\bibinfo {volume} {97}} (\bibinfo {year} {2018})}\BibitemShut {NoStop}%
\bibitem [{\citenamefont {Ba{\~n}uls}(2018)}]{Bauls2018TensorNA}%
  \BibitemOpen
  \bibfield  {author} {\bibinfo {author} {\bibfnamefont {M.~C.}\ \bibnamefont {Ba{\~n}uls}},\ }\bibfield  {title} {\bibinfo {title} {Tensor networks and their use for lattice gauge theories},\ }\href {https://api.semanticscholar.org/CorpusID:119403852} {\bibfield  {journal} {\bibinfo  {journal} {Proceedings of The 36th Annual International Symposium on Lattice Field Theory — PoS(LATTICE2018)}\ } (\bibinfo {year} {2018})}\BibitemShut {NoStop}%
\bibitem [{\citenamefont {Silvi}\ \emph {et~al.}(2014)\citenamefont {Silvi}, \citenamefont {Rico}, \citenamefont {Calarco},\ and\ \citenamefont {Montangero}}]{Silvi_2014}%
  \BibitemOpen
  \bibfield  {author} {\bibinfo {author} {\bibfnamefont {P.}~\bibnamefont {Silvi}}, \bibinfo {author} {\bibfnamefont {E.}~\bibnamefont {Rico}}, \bibinfo {author} {\bibfnamefont {T.}~\bibnamefont {Calarco}},\ and\ \bibinfo {author} {\bibfnamefont {S.}~\bibnamefont {Montangero}},\ }\bibfield  {title} {\bibinfo {title} {Lattice gauge tensor networks},\ }\href {https://doi.org/10.1088/1367-2630/16/10/103015} {\bibfield  {journal} {\bibinfo  {journal} {New Journal of Physics}\ }\textbf {\bibinfo {volume} {16}},\ \bibinfo {pages} {103015} (\bibinfo {year} {2014})}\BibitemShut {NoStop}%
\bibitem [{\citenamefont {Rico}\ \emph {et~al.}(2014)\citenamefont {Rico}, \citenamefont {Pichler}, \citenamefont {Dalmonte}, \citenamefont {Zoller},\ and\ \citenamefont {Montangero}}]{PhysRevLett.112.201601}%
  \BibitemOpen
  \bibfield  {author} {\bibinfo {author} {\bibfnamefont {E.}~\bibnamefont {Rico}}, \bibinfo {author} {\bibfnamefont {T.}~\bibnamefont {Pichler}}, \bibinfo {author} {\bibfnamefont {M.}~\bibnamefont {Dalmonte}}, \bibinfo {author} {\bibfnamefont {P.}~\bibnamefont {Zoller}},\ and\ \bibinfo {author} {\bibfnamefont {S.}~\bibnamefont {Montangero}},\ }\bibfield  {title} {\bibinfo {title} {Tensor networks for lattice gauge theories and atomic quantum simulation},\ }\href {https://doi.org/10.1103/PhysRevLett.112.201601} {\bibfield  {journal} {\bibinfo  {journal} {Phys. Rev. Lett.}\ }\textbf {\bibinfo {volume} {112}},\ \bibinfo {pages} {201601} (\bibinfo {year} {2014})}\BibitemShut {NoStop}%
\bibitem [{\citenamefont {Mathew}\ and\ \citenamefont {Raychowdhury}(2022)}]{Mathew_2022}%
  \BibitemOpen
  \bibfield  {author} {\bibinfo {author} {\bibfnamefont {E.}~\bibnamefont {Mathew}}\ and\ \bibinfo {author} {\bibfnamefont {I.}~\bibnamefont {Raychowdhury}},\ }\bibfield  {title} {\bibinfo {title} {Protecting local and global symmetries in simulating $z_2$ non-abelian gauge theories},\ }\bibfield  {journal} {\bibinfo  {journal} {Physical Review D}\ }\textbf {\bibinfo {volume} {106}},\ \href {https://doi.org/10.1103/physrevd.106.054510} {10.1103/physrevd.106.054510} (\bibinfo {year} {2022})\BibitemShut {NoStop}%
\bibitem [{\citenamefont {Mazzola}\ \emph {et~al.}(2021)\citenamefont {Mazzola}, \citenamefont {Mathis}, \citenamefont {Mazzola},\ and\ \citenamefont {Tavernelli}}]{Mazzola_2021}%
  \BibitemOpen
  \bibfield  {author} {\bibinfo {author} {\bibfnamefont {G.}~\bibnamefont {Mazzola}}, \bibinfo {author} {\bibfnamefont {S.~V.}\ \bibnamefont {Mathis}}, \bibinfo {author} {\bibfnamefont {G.}~\bibnamefont {Mazzola}},\ and\ \bibinfo {author} {\bibfnamefont {I.}~\bibnamefont {Tavernelli}},\ }\bibfield  {title} {\bibinfo {title} {Gauge-invariant quantum circuits for u(1) and yang-mills lattice gauge theories},\ }\bibfield  {journal} {\bibinfo  {journal} {Physical Review Research}\ }\textbf {\bibinfo {volume} {3}},\ \href {https://doi.org/10.1103/physrevresearch.3.043209} {10.1103/physrevresearch.3.043209} (\bibinfo {year} {2021})\BibitemShut {NoStop}%
\bibitem [{\citenamefont {Frank}\ \emph {et~al.}(2020)\citenamefont {Frank}, \citenamefont {Huffman},\ and\ \citenamefont {Chandrasekharan}}]{Frank_2020}%
  \BibitemOpen
  \bibfield  {author} {\bibinfo {author} {\bibfnamefont {J.}~\bibnamefont {Frank}}, \bibinfo {author} {\bibfnamefont {E.}~\bibnamefont {Huffman}},\ and\ \bibinfo {author} {\bibfnamefont {S.}~\bibnamefont {Chandrasekharan}},\ }\bibfield  {title} {\bibinfo {title} {Emergence of gauss’ law in a z2 lattice gauge theory in 1+1 dimensions},\ }\href {https://doi.org/10.1016/j.physletb.2020.135484} {\bibfield  {journal} {\bibinfo  {journal} {Physics Letters B}\ }\textbf {\bibinfo {volume} {806}},\ \bibinfo {pages} {135484} (\bibinfo {year} {2020})}\BibitemShut {NoStop}%
\bibitem [{\citenamefont {Davoudi}\ \emph {et~al.}(2023)\citenamefont {Davoudi}, \citenamefont {Mueller},\ and\ \citenamefont {Powers}}]{Davoudi-PRL-2023}%
  \BibitemOpen
  \bibfield  {author} {\bibinfo {author} {\bibfnamefont {Z.}~\bibnamefont {Davoudi}}, \bibinfo {author} {\bibfnamefont {N.}~\bibnamefont {Mueller}},\ and\ \bibinfo {author} {\bibfnamefont {C.}~\bibnamefont {Powers}},\ }\bibfield  {title} {\bibinfo {title} {Towards quantum computing phase diagrams of gauge theories with thermal pure quantum states},\ }\href {https://doi.org/10.1103/PhysRevLett.131.081901} {\bibfield  {journal} {\bibinfo  {journal} {Phys. Rev. Lett.}\ }\textbf {\bibinfo {volume} {131}},\ \bibinfo {pages} {081901} (\bibinfo {year} {2023})}\BibitemShut {NoStop}%
\bibitem [{\citenamefont {Kuroiwa}\ and\ \citenamefont {Nakagawa}(2021)}]{PhysRevResearch.3.013197}%
  \BibitemOpen
  \bibfield  {author} {\bibinfo {author} {\bibfnamefont {K.}~\bibnamefont {Kuroiwa}}\ and\ \bibinfo {author} {\bibfnamefont {Y.~O.}\ \bibnamefont {Nakagawa}},\ }\bibfield  {title} {\bibinfo {title} {Penalty methods for a variational quantum eigensolver},\ }\href {https://doi.org/10.1103/PhysRevResearch.3.013197} {\bibfield  {journal} {\bibinfo  {journal} {Phys. Rev. Res.}\ }\textbf {\bibinfo {volume} {3}},\ \bibinfo {pages} {013197} (\bibinfo {year} {2021})}\BibitemShut {NoStop}%
\bibitem [{\citenamefont {Ryabinkin}\ \emph {et~al.}(2018)\citenamefont {Ryabinkin}, \citenamefont {Genin},\ and\ \citenamefont {Izmaylov}}]{article5}%
  \BibitemOpen
  \bibfield  {author} {\bibinfo {author} {\bibfnamefont {I.}~\bibnamefont {Ryabinkin}}, \bibinfo {author} {\bibfnamefont {S.}~\bibnamefont {Genin}},\ and\ \bibinfo {author} {\bibfnamefont {A.}~\bibnamefont {Izmaylov}},\ }\bibfield  {title} {\bibinfo {title} {Constrained variational quantum eigensolver: Quantum computer search engine in the fock space},\ }\href {https://doi.org/10.1021/acs.jctc.8b00943} {\bibfield  {journal} {\bibinfo  {journal} {Journal of Chemical Theory and Computation}\ }\textbf {\bibinfo {volume} {15}} (\bibinfo {year} {2018})}\BibitemShut {NoStop}%
\bibitem [{\citenamefont {Caruana}(1997)}]{Caruana1997}%
  \BibitemOpen
  \bibfield  {author} {\bibinfo {author} {\bibfnamefont {R.}~\bibnamefont {Caruana}},\ }\bibfield  {title} {\bibinfo {title} {Multitask learning},\ }\href {https://doi.org/10.1023/A:1007379606734} {\bibfield  {journal} {\bibinfo  {journal} {Machine Learning}\ }\textbf {\bibinfo {volume} {28}},\ \bibinfo {pages} {015026} (\bibinfo {year} {1997})}\BibitemShut {NoStop}%
\bibitem [{\citenamefont {Zhang}\ and\ \citenamefont {Yang}(2022)}]{9392366}%
  \BibitemOpen
  \bibfield  {author} {\bibinfo {author} {\bibfnamefont {Y.}~\bibnamefont {Zhang}}\ and\ \bibinfo {author} {\bibfnamefont {Q.}~\bibnamefont {Yang}},\ }\bibfield  {title} {\bibinfo {title} {A survey on multi-task learning},\ }\href {https://doi.org/10.1109/TKDE.2021.3070203} {\bibfield  {journal} {\bibinfo  {journal} {IEEE Transactions on Knowledge and Data Engineering}\ }\textbf {\bibinfo {volume} {34}},\ \bibinfo {pages} {5586} (\bibinfo {year} {2022})}\BibitemShut {NoStop}%
\bibitem [{\citenamefont {Thung}\ and\ \citenamefont {Wee}(2018)}]{articleq}%
  \BibitemOpen
  \bibfield  {author} {\bibinfo {author} {\bibfnamefont {K.}~\bibnamefont {Thung}}\ and\ \bibinfo {author} {\bibfnamefont {C.-Y.}\ \bibnamefont {Wee}},\ }\bibfield  {title} {\bibinfo {title} {A brief review on multi-task learning},\ }\href {https://doi.org/10.1007/s11042-018-6463-x} {\bibfield  {journal} {\bibinfo  {journal} {Multimedia Tools and Applications}\ }\textbf {\bibinfo {volume} {77}} (\bibinfo {year} {2018})}\BibitemShut {NoStop}%
\bibitem [{\citenamefont {Sener}\ and\ \citenamefont {Koltun}(2018)}]{NEURIPS2018_432aca3a}%
  \BibitemOpen
  \bibfield  {author} {\bibinfo {author} {\bibfnamefont {O.}~\bibnamefont {Sener}}\ and\ \bibinfo {author} {\bibfnamefont {V.}~\bibnamefont {Koltun}},\ }\bibfield  {title} {\bibinfo {title} {Multi-task learning as multi-objective optimization},\ }in\ \href {https://proceedings.neurips.cc/paper_files/paper/2018/file/432aca3a1e345e339f35a30c8f65edce-Paper.pdf} {\emph {\bibinfo {booktitle} {Advances in Neural Information Processing Systems}}},\ Vol.~\bibinfo {volume} {31},\ \bibinfo {editor} {edited by\ \bibinfo {editor} {\bibfnamefont {S.}~\bibnamefont {Bengio}}, \bibinfo {editor} {\bibfnamefont {H.}~\bibnamefont {Wallach}}, \bibinfo {editor} {\bibfnamefont {H.}~\bibnamefont {Larochelle}}, \bibinfo {editor} {\bibfnamefont {K.}~\bibnamefont {Grauman}}, \bibinfo {editor} {\bibfnamefont {N.}~\bibnamefont {Cesa-Bianchi}},\ and\ \bibinfo {editor} {\bibfnamefont {R.}~\bibnamefont {Garnett}}}\ (\bibinfo  {publisher} {Curran Associates, Inc.},\ \bibinfo {year} {2018})\BibitemShut {NoStop}%
\bibitem [{\citenamefont {Kuhn}\ and\ \citenamefont {Tucker}(2014)}]{Kuhn2014}%
  \BibitemOpen
  \bibfield  {author} {\bibinfo {author} {\bibfnamefont {H.~W.}\ \bibnamefont {Kuhn}}\ and\ \bibinfo {author} {\bibfnamefont {A.~W.}\ \bibnamefont {Tucker}},\ }\bibinfo {title} {Nonlinear programming},\ in\ \href {https://doi.org/10.1007/978-3-0348-0439-4_11} {\emph {\bibinfo {booktitle} {Traces and Emergence of Nonlinear Programming}}},\ \bibinfo {editor} {edited by\ \bibinfo {editor} {\bibfnamefont {G.}~\bibnamefont {Giorgi}}\ and\ \bibinfo {editor} {\bibfnamefont {T.~H.}\ \bibnamefont {Kjeldsen}}}\ (\bibinfo  {publisher} {Springer Basel},\ \bibinfo {address} {Basel},\ \bibinfo {year} {2014})\ pp.\ \bibinfo {pages} {247--258}\BibitemShut {NoStop}%
\bibitem [{\citenamefont {Désidéri}(2012)}]{DESIDERI2012313}%
  \BibitemOpen
  \bibfield  {author} {\bibinfo {author} {\bibfnamefont {J.-A.}\ \bibnamefont {Désidéri}},\ }\bibfield  {title} {\bibinfo {title} {Multiple-gradient descent algorithm (mgda) for multiobjective optimization},\ }\href {https://doi.org/https://doi.org/10.1016/j.crma.2012.03.014} {\bibfield  {journal} {\bibinfo  {journal} {Comptes Rendus Mathematique}\ }\textbf {\bibinfo {volume} {350}},\ \bibinfo {pages} {313} (\bibinfo {year} {2012})}\BibitemShut {NoStop}%
\bibitem [{\citenamefont {Jaggi}(2013)}]{pmlr-v28-jaggi13}%
  \BibitemOpen
  \bibfield  {author} {\bibinfo {author} {\bibfnamefont {M.}~\bibnamefont {Jaggi}},\ }\bibfield  {title} {\bibinfo {title} {Revisiting {Frank-Wolfe}: Projection-free sparse convex optimization},\ }in\ \href {https://proceedings.mlr.press/v28/jaggi13.html} {\emph {\bibinfo {booktitle} {Proceedings of the 30th International Conference on Machine Learning}}},\ \bibinfo {series} {Proceedings of Machine Learning Research}, Vol.~\bibinfo {volume} {28},\ \bibinfo {editor} {edited by\ \bibinfo {editor} {\bibfnamefont {S.}~\bibnamefont {Dasgupta}}\ and\ \bibinfo {editor} {\bibfnamefont {D.}~\bibnamefont {McAllester}}}\ (\bibinfo  {publisher} {PMLR},\ \bibinfo {address} {Atlanta, Georgia, USA},\ \bibinfo {year} {2013})\ pp.\ \bibinfo {pages} {427--435}\BibitemShut {NoStop}%
\bibitem [{\citenamefont {Borla}\ \emph {et~al.}(2020)\citenamefont {Borla}, \citenamefont {Verresen}, \citenamefont {Grusdt},\ and\ \citenamefont {Moroz}}]{Borla_2020}%
  \BibitemOpen
  \bibfield  {author} {\bibinfo {author} {\bibfnamefont {U.}~\bibnamefont {Borla}}, \bibinfo {author} {\bibfnamefont {R.}~\bibnamefont {Verresen}}, \bibinfo {author} {\bibfnamefont {F.}~\bibnamefont {Grusdt}},\ and\ \bibinfo {author} {\bibfnamefont {S.}~\bibnamefont {Moroz}},\ }\bibfield  {title} {\bibinfo {title} {Confined phases of one-dimensional spinless fermions coupled to z2 gauge theory},\ }\bibfield  {journal} {\bibinfo  {journal} {Physical Review Letters}\ }\textbf {\bibinfo {volume} {124}},\ \href {https://doi.org/10.1103/physrevlett.124.120503} {10.1103/physrevlett.124.120503} (\bibinfo {year} {2020})\BibitemShut {NoStop}%
\bibitem [{\citenamefont {Weinberg}\ and\ \citenamefont {Bukov}(2017)}]{10.21468/SciPostPhys.2.1.003}%
  \BibitemOpen
  \bibfield  {author} {\bibinfo {author} {\bibfnamefont {P.}~\bibnamefont {Weinberg}}\ and\ \bibinfo {author} {\bibfnamefont {M.}~\bibnamefont {Bukov}},\ }\bibfield  {title} {\bibinfo {title} {{QuSpin: a Python package for dynamics and exact diagonalisation of quantum many body systems part I: spin chains}},\ }\href {https://doi.org/10.21468/SciPostPhys.2.1.003} {\bibfield  {journal} {\bibinfo  {journal} {SciPost Phys.}\ }\textbf {\bibinfo {volume} {2}},\ \bibinfo {pages} {003} (\bibinfo {year} {2017})}\BibitemShut {NoStop}%
\bibitem [{\citenamefont {Weinberg}\ and\ \citenamefont {Bukov}(2019)}]{10.21468/SciPostPhys.7.2.020}%
  \BibitemOpen
  \bibfield  {author} {\bibinfo {author} {\bibfnamefont {P.}~\bibnamefont {Weinberg}}\ and\ \bibinfo {author} {\bibfnamefont {M.}~\bibnamefont {Bukov}},\ }\bibfield  {title} {\bibinfo {title} {{QuSpin: a Python package for dynamics and exact diagonalisation of quantum many body systems. Part II: bosons, fermions and higher spins}},\ }\href {https://doi.org/10.21468/SciPostPhys.7.2.020} {\bibfield  {journal} {\bibinfo  {journal} {SciPost Phys.}\ }\textbf {\bibinfo {volume} {7}},\ \bibinfo {pages} {020} (\bibinfo {year} {2019})}\BibitemShut {NoStop}%
\bibitem [{\citenamefont {Di~Pillo}\ and\ \citenamefont {Grippo}(1989)}]{doi:10.1137/0327068}%
  \BibitemOpen
  \bibfield  {author} {\bibinfo {author} {\bibfnamefont {G.}~\bibnamefont {Di~Pillo}}\ and\ \bibinfo {author} {\bibfnamefont {L.}~\bibnamefont {Grippo}},\ }\bibfield  {title} {\bibinfo {title} {Exact penalty functions in constrained optimization},\ }\href {https://doi.org/10.1137/0327068} {\bibfield  {journal} {\bibinfo  {journal} {SIAM Journal on Control and Optimization}\ }\textbf {\bibinfo {volume} {27}},\ \bibinfo {pages} {1333} (\bibinfo {year} {1989})},\ \Eprint {https://arxiv.org/abs/https://doi.org/10.1137/0327068} {https://doi.org/10.1137/0327068} \BibitemShut {NoStop}%
\bibitem [{\citenamefont {Virtanen}\ \emph {et~al.}(2020)\citenamefont {Virtanen}, \citenamefont {Gommers}, \citenamefont {Oliphant}, \citenamefont {Haberland}, \citenamefont {Reddy}, \citenamefont {Cournapeau}, \citenamefont {Burovski}, \citenamefont {Peterson}, \citenamefont {Weckesser}, \citenamefont {Bright}, \citenamefont {{van der Walt}}, \citenamefont {Brett}, \citenamefont {Wilson}, \citenamefont {Millman}, \citenamefont {Mayorov}, \citenamefont {Nelson}, \citenamefont {Jones}, \citenamefont {Kern}, \citenamefont {Larson}, \citenamefont {Carey}, \citenamefont {Polat}, \citenamefont {Feng}, \citenamefont {Moore}, \citenamefont {{VanderPlas}}, \citenamefont {Laxalde}, \citenamefont {Perktold}, \citenamefont {Cimrman}, \citenamefont {Henriksen}, \citenamefont {Quintero}, \citenamefont {Harris}, \citenamefont {Archibald}, \citenamefont {Ribeiro}, \citenamefont {Pedregosa}, \citenamefont {{van Mulbregt}},\ and\ \citenamefont {{SciPy 1.0 Contributors}}}]{2020SciPy-NMeth}%
  \BibitemOpen
  \bibfield  {author} {\bibinfo {author} {\bibfnamefont {P.}~\bibnamefont {Virtanen}}, \bibinfo {author} {\bibfnamefont {R.}~\bibnamefont {Gommers}}, \bibinfo {author} {\bibfnamefont {T.~E.}\ \bibnamefont {Oliphant}}, \bibinfo {author} {\bibfnamefont {M.}~\bibnamefont {Haberland}}, \bibinfo {author} {\bibfnamefont {T.}~\bibnamefont {Reddy}}, \bibinfo {author} {\bibfnamefont {D.}~\bibnamefont {Cournapeau}}, \bibinfo {author} {\bibfnamefont {E.}~\bibnamefont {Burovski}}, \bibinfo {author} {\bibfnamefont {P.}~\bibnamefont {Peterson}}, \bibinfo {author} {\bibfnamefont {W.}~\bibnamefont {Weckesser}}, \bibinfo {author} {\bibfnamefont {J.}~\bibnamefont {Bright}}, \bibinfo {author} {\bibfnamefont {S.~J.}\ \bibnamefont {{van der Walt}}}, \bibinfo {author} {\bibfnamefont {M.}~\bibnamefont {Brett}}, \bibinfo {author} {\bibfnamefont {J.}~\bibnamefont {Wilson}}, \bibinfo {author} {\bibfnamefont {K.~J.}\ \bibnamefont {Millman}}, \bibinfo {author} {\bibfnamefont {N.}~\bibnamefont {Mayorov}}, \bibinfo {author} {\bibfnamefont
  {A.~R.~J.}\ \bibnamefont {Nelson}}, \bibinfo {author} {\bibfnamefont {E.}~\bibnamefont {Jones}}, \bibinfo {author} {\bibfnamefont {R.}~\bibnamefont {Kern}}, \bibinfo {author} {\bibfnamefont {E.}~\bibnamefont {Larson}}, \bibinfo {author} {\bibfnamefont {C.~J.}\ \bibnamefont {Carey}}, \bibinfo {author} {\bibfnamefont {{\.I}.}~\bibnamefont {Polat}}, \bibinfo {author} {\bibfnamefont {Y.}~\bibnamefont {Feng}}, \bibinfo {author} {\bibfnamefont {E.~W.}\ \bibnamefont {Moore}}, \bibinfo {author} {\bibfnamefont {J.}~\bibnamefont {{VanderPlas}}}, \bibinfo {author} {\bibfnamefont {D.}~\bibnamefont {Laxalde}}, \bibinfo {author} {\bibfnamefont {J.}~\bibnamefont {Perktold}}, \bibinfo {author} {\bibfnamefont {R.}~\bibnamefont {Cimrman}}, \bibinfo {author} {\bibfnamefont {I.}~\bibnamefont {Henriksen}}, \bibinfo {author} {\bibfnamefont {E.~A.}\ \bibnamefont {Quintero}}, \bibinfo {author} {\bibfnamefont {C.~R.}\ \bibnamefont {Harris}}, \bibinfo {author} {\bibfnamefont {A.~M.}\ \bibnamefont {Archibald}}, \bibinfo {author}
  {\bibfnamefont {A.~H.}\ \bibnamefont {Ribeiro}}, \bibinfo {author} {\bibfnamefont {F.}~\bibnamefont {Pedregosa}}, \bibinfo {author} {\bibfnamefont {P.}~\bibnamefont {{van Mulbregt}}},\ and\ \bibinfo {author} {\bibnamefont {{SciPy 1.0 Contributors}}},\ }\bibfield  {title} {\bibinfo {title} {{{SciPy} 1.0: Fundamental Algorithms for Scientific Computing in Python}},\ }\href {https://doi.org/10.1038/s41592-019-0686-2} {\bibfield  {journal} {\bibinfo  {journal} {Nature Methods}\ }\textbf {\bibinfo {volume} {17}},\ \bibinfo {pages} {261} (\bibinfo {year} {2020})}\BibitemShut {NoStop}%
\end{thebibliography}%
\end{document}